\documentclass[twocolumn]{revtex4}
\usepackage{graphicx}
\usepackage[]{epsfig}

\newcommand{\beq}{\begin{equation}}
\newcommand{\eeq}{\end{equation}}
\newcommand{\beqd}{\begin{displaymath}}
\newcommand{\eeqd}{\end{displaymath}}
\newcommand{\beqa}{\begin{eqnarray}}
\newcommand{\eeqa}{\end{eqnarray}}
\newcommand{\bew}{\begin{widetext}}
\newcommand{\eew}{\end{widetext}}

\newcommand{\sign}{{\rm sign}}

\newcommand{\comment}[1]{}

\begin{document}
\title{Dynamical Landau Theory of the Glass Crossover}

\author{Tommaso Rizzo$^{1,2}$}
\affiliation{
$^1$ ISC-CNR, UOS Rome, Universit\`a "Sapienza", PIazzale A. Moro 2, \\
$^2$ Dip. Fisica, Universit\`a "Sapienza", Piazzale A. Moro 2, I-00185, Rome, Italy \\
I-00185, Rome, Italy}

\begin{abstract}

I introduce a dynamical field theory to describe the glassy behavior in supercooled liquids.
The mean-field approximation of the theory predicts a dynamical arrest transition, as in ideal Mode-Coupling-Theory and mean-field discontinuous Spin-Glass Models. Instead {\it beyond} the mean-field approximation the theory predicts that the transition is avoided and transformed into a crossover, as observed in experiments and simulations.
To go beyond mean-field a standard perturbative loop expansion is performed at first. Approaching the ideal critical point this expansion is divergent at all orders and I show that the leading divergent term at any given order is the same of a dynamical stochastic equation, called Stochastic-Beta-Relaxation (SBR) in {\it EPL 106, 56003 (2014)}.
At variance with the original theory SBR can be studied beyond mean-field directly, without the need to resort to a perturbative expansion. Thus it provides a qualitative and quantitative description of the dynamical crossover.
For consistency reasons it is important to establish the connection between the dynamical field theory and  SBR beyond perturbation theory.
This can be done with the help of a stronger result: the dynamical field theory is {\it exactly} equivalent to a theory with quenched disorder. Qualitatively the non-perturbative mechanism leading to the crossover is therefore the same of SBR. Quantitatively SBR corresponds to make the mean-field approximation once the quenched disorder has been generated.
\end{abstract}

\maketitle

\section{Introduction}

Ideal Mode-Coupling-Theory (MCT) for supercooled liquids \cite{Gotze09} and mean-field discontinuous Spin-Glass (SG) Models \cite{Kirkpatrick87c,Kirkpatrick87d,Crisanti93} both predict dynamical arrest at a critical temperature.
Experiments and numerical simulations on the other hand do not display a transition but rather a dynamical crossover from relaxational to activated-like dynamics.
In a recent publication \cite{Rizzo14} I have introduced a dynamical stochastic equation, called Stochastic-Beta-Relaxation (SBR), that provides a characterization of the temperature region where the crossover occurs.

As a model of the glass crossover SBR is rather consistent and lack many of the drawbacks of earlier proposal to amend ideal MCT. 
Ongoing studies of SBR are  unveiling a rich phenomenology and a rather non-trivial characterization of the qualitative and quantitative changes occurring at the crossover \cite{Rizzo15a,Rizzo15b}. These  include notably a change in the spatial structure of dynamical fluctuations characterized by the appearance of strong Dynamical Heterogeneities and violations of the Stokes-Einstein relationship.
Additionally the increase of the relaxation time and of the dynamic susceptibility is accompanied by a {\it decrease} of the dynamical correlation length below the crossover temperature \cite{Rizzo15b}. This challenges somehow the classic Adams and Gibbs \cite{Adams65} view that dynamical slowing down is essentially driven by the increase of a correlation length associated to the size of the Cooperatively Rearranging Regions.

In order to fully appreciate the aforementioned properties one should acknowledge that SBR is not a phenomenological theory. SBR is characterized by random quenched fluctuations of the temperature. This essential feature is not an {\it ad hoc} hypothesis, instead it comes out from a rigorous and complex computation starting from clear and falsifiable assumptions. 
As such SBR stands in a different position compared to many theories in the field of glassy systems that are either explicitly phenomenological or at least very speculative. 
In this paper I present a complete discussion of this issue.

The starting point of the argument is that both MCT  and discontinuous SG (characterized by one step of Parisi's Replica-Symmetry-Breaking (1RSB) \cite{MPV}) obey G{\"o}tze's equation for the critical correlator. 
This is a mean-field equation in the sense that it is a polynomial (quadratic) equation with coefficients that are regular functions of the external parameters (temperature, pressure, magnetic field, etc.). 
Landau theory suggests that if the equation holds {\it approaching} a critical point then the critical point itself should be described by an effective Hamiltonian.

The effective Hamiltonian (or effective action/theory in a broader dynamical context) is simpler that the original microscopic theory and can be identified by means of symmetry arguments.
As a first guess the effective theory can be determined by integration of the mean-field equation. This guarantees that the effective theory is extremized by the solutions of the mean-field equation. More precisely the effective theory is the most general polynomial of the order parameter of the required order (cubic in our case, because G{\"o}tze's equation is quadratic) with the symmetries of the original microscopic problem \footnote{This criterion is mandatory if, as in this case, there are terms that are irrelevant at the level of the mean-field equation and would not show up upon integration}.    

The above arguments are developed in section \ref{GCT}. In the first subsection I introduce a field-theoretical formulation of the dynamics in order to understand the properties and symmetries of the order parameter and of the action. Here I discuss the so-called fast-motion (FM) limit. In this limit, dynamical correlations, and thus the effective theory, acquire a Replica-Symmetric(RS)-like structure. In the second subsection I discuss in more details the nature of dynamical arrest in MCT and 1RSB-SG. I recall the standard arguments of Landau theory and the need to replace the critical mean-field equations with an effective theory. I then argue that the effective theory is RS-like in the FM limit but it retains this structure also in the $\beta$-regime, where the correlations are not RS-like. 
In the last subsection the dynamical effective theory is presented, for brevity I refer to this specific effective action as the Glassy Critical Theory (GCT).

Section \ref{GCT} is essentially introductory and deal mostly with already published work.
The following sections are the body of the paper and are devoted to the study of the GCT.
In section \ref{sloop} I study the GCT by means of a loop expansion. I start with the mean-field equation, exhibit the bare propagator and then study the Feynman diagrams. As usual, each term in the loop expansion is found to be divergent approaching the ideal critical point. I show that the leading divergent term at any given order is the same generated by the loop expansion of SBR. This proves a statement that was condensed in few lines in \cite{Rizzo14}.
The discussion is carried on initially for the zero-dimensional case, in the last subsection it is generalized to finite-dimensional systems. 
I note that the mapping to SBR will be proven for all correlation functions while in \cite{Rizzo14} only the average order parameter was mentioned.

At variance with the GCT, SBR can be studied non-perturbatively unveiling the avoided nature of the transition and yielding a description of the dynamical crossover.
It is therefore important to establish the connection between the dynamical field theory and  SBR beyond perturbation theory.
This can be done with the help of a result, presented in section \ref{beyond}, that is stronger than the one presented in \cite{Rizzo14}. According to it the dynamical field theory is {\it exactly} equivalent to a dynamical theory with quenched disorder. Qualitatively the non-perturbative mechanism leading to the crossover is therefore the same of SBR. Quantitatively SBR corresponds to make the mean-field approximation once the quenched disorder has been generated and the quality of this approximation depends on the value of a dimensionless constant.

A well-known prediction within the Random-First-Order-Transition (RFOT) theory of the glass transition \cite{Kirkpatrick87d} is that glassyness is driven by the increase of the correlation length upon lowering the temperature. By contrast, SBR suggests a different scenario at least at the crossover. The interesting point is that the alerted reader will recognize that the GCT, albeit being a dynamical theory, is closely related to the static replica theory that describes the so-called dynamical transition within RFOT. Therefore it seems that there is a contradiction that need to be resolved, either by reconciling the two pictures or by ruling out one of them.

Summarizing, section \ref{GCT} is where I lay down the basic assumption {\it a la} Landau, {\it i.e.} the fact at the dynamical crossover G{\"o}tze's critical equation must be replaced by the Glassy Critical Theory (GCT) defined in eq. (\ref{action}). Section \ref{sloop} and \ref{beyond} present instead rigorous results that follow from the assumption. In section \ref{Conclusions} I give my conclusions.

\section{Glassy Critical Theory}
\label{GCT}

\subsection{Dynamical Field Theory}

In order to determine the effective theory for the glass crossover I introduce a field-theoretical representation of dynamics.
I illustrate the method in the simplest case of a single variable $q$ that obeys Langevin dynamics :
\beq
{1 \over \Gamma_0}\dot{q}=-\beta {d H \over dq}+\xi\, ,\  \langle \xi(t)\xi(t')\rangle={2 \over \Gamma_0}\delta(t-t') \ .
\eeq
The discussion can be easily generalized to systems with many degrees of freedom.
In the standard framework \cite{Martin78,DeDominicis78,Kurchan92,Zinn02} one typically sets the initial time of dynamics to minus infinity,  here instead I want to compute averages of observables $O[q(t)]$ starting from an equilibrium initial configuration at time $t=0$:
\begin{widetext}
\beq
\langle O \rangle \equiv {1 \over \tilde{Z}}\int \prod \left[dq(t) dG(\xi(t)) \delta \left(-{1 \over \Gamma_0}\dot{q}-\beta {d H \over dq}+\xi\right)\right] O \exp[-\beta H(q(0))]
\eeq
\end{widetext}
where $dG(\xi(t))$ is the Gaussian measure over the thermal noise $\eta$ \footnote{Even if the argument of the $\delta$ function is the Langevin equation, it is also a $\delta$ function on its solution because the Jacobian is one if we discretize in time according to the Ito convention.}. The average over the initial condition implies that the normalization constant $\tilde{Z}$ is the exactly the partition function.
Rewriting the delta function in exponential form by means of an auxiliary variable $\hat{q}(t)$ and integrating over $\xi$ we have:
\beq
\langle O \rangle =
{1 \over Z}\int \prod[dq(t)d \hat{q}(t)] O \exp[-{\mathcal L }]
\eeq
where
\beq
{\mathcal L}=\int dt \left[{1 \over \Gamma_0}\left(\hat{q}\dot{q}-\hat{q}^2\right) +\hat{q} \beta {d H \over dq}\right]+\beta H(q(0))
\label{Lsingle}
\eeq
The normalization constant $Z$ is proportional to the partition function at initial time (times an irrelevant  numerical constant).
Averages of observables involving the variable $\hat{q}$ are associated to the response to perturbations of the form $\Delta H(t)= -h(t)q(t)$ for times {\it larger} than the initial condition ($t>0$).  For instance it can be easily checked that:
\beq
R(t,t')\equiv {\delta \langle q(t) \rangle \over \delta h(t')}=\beta\langle q(t) \hat{q}(t') \rangle  \ \, t'>0
\label{RRt}
\eeq
Responses to perturbation of the initial condition obey the standard Fluctuation-Dissipation-Theorem  
\beq
R(t)\equiv {d \langle q(t) \rangle \over d h(0)}=\beta\langle q(t)q(0)\rangle_c
\label{RR0}
\eeq
For a generic distribution of the initial condition, causality implies that
\beq
R(t,t')=0\,, \ t'>t \ .
\eeq
Since the initial condition is equilibrated, correlation functions are time-translational invariant (TTI):
\beq
\langle q(t)q(t')\rangle=C(t-t')
\eeq
where $C(s)=C(-s)$.
Furthermore the Fluctuation-Dissipation-Theorem (FDT) implies that the response to  a perturbation acting on the initial condition and at all later times up to $t''<t$ is given by:
\beq
R(t)+\int_{0^+}^{t''} R(t,t') dt'=\beta\langle q(t)q(t'')\rangle_c\ .
\eeq
The above relationships combined imply:
\beq
 \langle q(t) \hat{q}(t') \rangle ={d \over dt'}\langle q(t')q(t)\rangle=-\dot{C}(t-t') \ \  (t'<t) \ .
\label{qhq}
\eeq
General relationship (valid also for off-equilibrium initial conditions) in the above formalism are also: 
\beq
 \langle\hat{q}(t) \rangle =0\ ,  \langle\hat{q}(t)\hat{q}(t') \rangle =0 \ .
\label{hqhq}
\eeq
They can be proven computing the derivative of $\ln \int \exp[-{\mathcal L}]$ with respect to fields $h(t)$ and noticing that $\ln \int \exp[-{\mathcal L}]=\ln Z$ (the free energy at the initial time) depends only on $h(0)$.

The part of the Lagrangian that depends explicitly on the Hamiltonian can be written in a compact form introducing a novel cordinate $\eta$ that has the properties of the product of two Grassmanian variables:
\beq
\int d\eta=0\, , \int \eta \, d\eta=1\,, \eta^2=0
\eeq   
 The variables $q(t)$ and $\hat{q}(t)$ can then be combined into a single variable parameterized by the coordinate $a=(t_a,\eta_a)$:
 \beq
q(a)\equiv q(t_a)+\eta_a \hat{q}(t_a) \ .
 \eeq
Given a function $A(a) \equiv A(t_a,\eta_a)$, I define the integral over $da$ as:
\beq
\int A(a) da \equiv \int_{0^+}^{t_{max}} A(t_a,\eta_a) d\eta_a dt_a+A(0,0)
\label{intdef}
\eeq
where $t_{max}$ is some maximal time over which we study dynamics whose precise value is irrelevant as long as it is larger than the larger time separation we want to study. 
We can write the action as the sum of a kinetic term proportional to $1/\Gamma_0$ and of a potential term:
\beq
{\mathcal L}={1 \over \Gamma_0}\int dt \left(\hat{q}\dot{q}-\hat{q}^2\right)+\int da \, \beta \, H(q(a))\ .
\label{Ldec}
\eeq
Note that the generic product $q(a)q(b)$ has four independent components but the various relationships derived above imply that instead the components of its average can be expressed in terms of the sole correlation function $C(s)$:
\begin{widetext}
\beqa
\langle q(a)q(b) \rangle &  = & \langle q(t_a)q(t_b)\rangle +\langle q(t_a)\hat{q}(t_b)\rangle \eta_b+\langle q(t_b)\hat{q}(t_a)\rangle \eta_a+\langle \hat{q}(t_b)\hat{q}(t_a)\rangle \eta_a \eta_b=
\nonumber
\\
& = & C(|t_a-t_b|)-\dot{C}(t_a-t_b) \theta(t_a-t_b) \eta_b-\dot{C}(t_b-t_a) \theta(t_b-t_a) \eta_a
\label{2time}
\eeqa
\end{widetext}
The above structure is very general: considering any two observables  $O_1(q)$, $O_2(q)$ one can show that the correlation $\langle O_1(q(a)) O_2(q(b))\rangle$ takes the form of the r.h.s. of (\ref{2time}) with $C(t)$ replaced by the correlation $C_{O_1O_2}(t)$.

I now introduce the so-called fast motion (FM) limit \cite{Kurchan92,Parisi13}. In physical terms it amounts to study relaxational dynamics on an  infinitely large timescale so that any two-time correlation $\langle q(t)q(t')\rangle$ has relaxed to its infinite-time value $\lim_{t \rightarrow \infty} \langle q(0)q(t)\rangle$. Formally this can be achieved by sending the parameter $\Gamma_0$ to infinity leading to an extremely fast microscopic dynamics (hence the name fast motion) so that the memory of the initial condition is lost instantaneously.
One can argue that in the FM limit the time dependence of any two-time correlation is considerably simpler, the correlation can take just two values depending if the two times are equal or different.
One can check by a simple explicit computation that the form (\ref{2time}) becomes:
\beq
\langle q(a) q(b)\rangle=\delta(ab)(C(0)-C(\infty))+C(\infty)
\label{2RS}
\eeq
where the delta function is defined with respect to the integral (\ref{intdef}) as $\int \delta(ab)g(a)da =g(b)$ and takes the form $\delta(ab)=\delta(t_a-t_b)(\eta_a+\eta_b)\mathcal{Q}_0(t_a)\mathcal{Q}_0(t_b)+\mathcal{P}_0(t_a)\mathcal{P}_0(t_b)$ where the function $\mathcal{P}_0(t)$ is defined as being equal to one if $t=0$ and zero otherwise. The function $\mathcal{Q}_0(t)$ is equal to $1-\mathcal{P}_0(t)$ \footnote{The unusual appearance of the functions $\mathcal{P}_0(t)$ and $\mathcal{Q}_0(t)$ is a consequence of the special role of the initial time $t=0$ encoded in the definition (\ref{intdef}). The explicit computation shows that $\int \delta(ab)g(a)da$ is equal to $g(b)=g(t_b)+\eta_b \hat{g}(t_b)$ for $t_b> 0$. For $t_b=0$ the integral gives just $g(0)$ which differs from $\neq g(0)+\hat{g}(0)\eta_a$. Nevertheless the identity holds {\it almost everywhere} because  upon an additional integration of the form (\ref{intdef})  $\hat{g}(0)$ has zero measure, while $g(0)$ has measure one. Therefore one can replace a two-time correlation in the FM limit with a delta function whenever integrations over all indexes are performed.}. 
Higher orders correlations also have a simplified structure, for instance a three-point correlation can take three possible values and takes the form:
\beq
\langle q(a) q(b) q(c)\rangle=c_0+c_1[\delta(ab)+\delta(ac)+\delta(cb)]+c_2 \delta(ab)\delta(ac)
\label{3RS}
\eeq
In general the value of an $n$-point correlation depends only on whether some indexes coincide. This implies the following general result: correlations in the FM limit have the same expression of replica-symmetric correlations in a  system of $n$ replicas, where $a,b=1,\dots n$, and $\delta(ab)$ is the  Kronecker delta.
I note that using this property one can establish the connection between dynamics and statics. In particular for solvable models like mean-field SG the statics can be solved by the replica method and one can recover the same results within dynamics by taking the FM limit \cite{Kurchan92,Parisi13}.

The connection between FM and replica symmetry follows from rather general arguments and holds also when using Newtonian dynamics with Hamilton's equations of motion.
In this case the only averaging occurs for the initial condition, more precisely once the initial configuration of the $q$ is fixed, the Maxwell-Boltzmann average over the momenta $p$ at the initial time generate different trajectories in analogy with the different thermal histories in Langevin dynamics. 
Let us introduce a field-theoretical representation of Hamiltonian dynamics considering once again a single degree of freedom $q$ and its conjugate momentum $p$. The action (\ref{Lsingle}) becomes for an Hamiltonian $H(q,p)$:
\beq
{\mathcal L}=\int dt \left[{\beta \over \Gamma_0}(\hat{q}\dot{p}-\hat{p}\dot{q}) +\hat{q} \beta {d H \over dq}+\hat{p} \beta {d H \over dp}\right]+\beta H(q(0),p(0))
\eeq
where $\Gamma_0=1$ but can take any value if we rescale times.
Using the same definitions for $a$ and $\int da$ introduced above, we can define combined variables:
\beq
q(a)\equiv q(t_a)+\eta_a \hat{q}(t_a) \ , p(a)\equiv p(t_a)+\eta_a \hat{p}(t_a)\, ,
\eeq
and again decompose the dynamical action into a kinetic and a potential-like term:
\beq
{\mathcal L}={\beta \over \Gamma_0}\int dt (\hat{q}\dot{p}-\hat{p}\dot{q})+\int da \, \beta \, H(q(a),p(a))
\eeq
One can check that relationships (\ref{RRt},  \ref{RR0},\ref{qhq},\ref{hqhq}) hold in this formalism too and so does (\ref{2time}). The FM limit corresponds to consider very large times scale or equivalently to have very fast microscopic dynamics {\it i.e.} sending $\Gamma_0$ to infinity and  again one can check that all dynamical correlations are RS-like.

\subsection{Glassy Phase Transitions}

Both mean-field 1RSB-SG  and MCT are characterized by phase transitions from a liquid phase to a glassy phase, the order parameter being a two-time correlation.
Although the order parameter is discontinuous, overall the transition is second-order in nature, notably: i) the correlator displays a square-root singularity, ii) the correlation length diverges and iii) dynamical fluctuations are also divergent. 
All these features can be shown explicitly in 1RSB-SG \cite{Kirkpatrick87c,Crisanti93}. In both cases the correlations obey the so-called G{\"o}tze's equation for the critical correlator that displays a square root singularity. Instead, fluctuations are not directly accessible in MCT but the presence of a diverging correlation length  can be demonstrated within Inhomogeneous MCT (IMCT) {\it i.e.} studying the response to spatially inhomogeneous external parameters \cite{Biroli06}.

MCT and 1RSB-SG are mean-field in the sense that the average order parameter near the critical temperature obeys a polynomial equation with coefficients that are regular functions of the external parameter. To illustrate this point let us consider the glassy phase. Here things are simpler because one can forget about dynamics and study the long time limit of the correlator. One finds
 that it obeys a quadratic equation of the form $\tau+q^2=0$  where $\tau$ vanishes linearly at the critical temperature and $q$ is the difference with the value of the correlator at $\tau=0$. The equation leads to the square-root singularity of the solution and is completely general. In particular it does not depend on the precise nature of the approximations we made: whenever we deal with an approximate closed equation for the averaged order parameter we will obtain the square-root singularity, {\it i.e.} mean-field behavior.
 
According to the modern theory of phase transitions the mean-field critical exponents can be altered by long-wavelength fluctuations below the upper critical dimension.
This implies that every mean-field prediction (including those of MCT) should be tested against these fluctuations. Within Landau theory one can argue that this problem can be studied considering an effective Hamiltonian/action for the long-wavelength fluctuations that depends on few coupling constants.  The effective action is a simple polynomial of the order parameter with the symmetries of the original problem.

Let us apply the previous consideration to 1RSB-SG and supercooled liquids to determine what is the Landau theory describing their critical point.
In general the long-wavelength order parameter is defined in terms of the Fourier transform of the microscopic order parameter at small momenta.
In Spin-Glasses one typically considers the Fourier transform of the microscopic overlap at different times:
\beq
Q(p,t,t')\equiv \sum_x s_x(t)s_x(t') e^{i p x}\ .
\eeq
In supercooled liquids the natural order parameter is  the density-density correlator at different times defined as 
\beq
\Phi(p,\Delta,t,t') \equiv \int d^Dx  \delta \rho(x,t)\delta \rho(x+\Delta,t) e^{i p x}
\eeq
where the density fluctuations are defined as $\delta \rho(x,t)\equiv \sum_i \delta(x-x_i(t))-\rho$ where $x_i(t)$ labels the position of particle $i$ at time $t$ and $\rho$ is the density.
The space coordinate $x$ lies on a lattice for SG systems while it is continuous in liquids. 

In both mean-field 1RSB-SG  and MCT near the critical temperature the correlator is concentrated near a system-dependent plateau value on the time-scale of the $\beta$ regime.
The deviations from the plateau value in the $\beta$ regime are small and critical and they should be identified with the critical order parameter.
Within MCT (but the same scenario is obtained within other treatments \cite{Franz12}) the critical order parameter is actually the projection of the distance from the plateau on a system-dependent critical mode $\tilde{H}_\Delta$, we thus have:
\beqa
 Q(p,t,t') & \approx &  q \delta(p)+g(p,t,t')\ ,
 \nonumber
 \\
   \Phi(p,\Delta,t,t') & \approx & F_\Delta \delta(p)+\tilde{H}_\Delta \,g(p,t,t') 
\label{cOP}
\eeqa
and the problem is to determine the effective theory obeyed by $g(p,t,t')$ in order to compute the average order parameter:
\beqa
 \langle Q(p,t,t') \rangle & \approx &  q \delta(p)+\langle g(p,t,t') \rangle  \ \ ,
\nonumber
\\ 
   \langle \Phi(p,\Delta,t,t') \rangle & \approx &  \tilde{F}_\Delta \delta(p)+\tilde{H}_\Delta \,\langle g(p,t,t') \rangle 
\eeqa
Fourier transform of the previous expression with respect to the real space variable $\Delta$ leads to the MCT expression
\beq
\langle \delta \rho_k(t)\delta \rho_{-k+p}(t') \rangle \approx F_k \delta(p)+H_k \,\langle g(p,t,t') \rangle \ .
\eeq
From higher order powers of (\ref{cOP}) one obtains that
higher orders connected correlations are also expressed near the critical temperature as averages of the critical field
\beq
\langle Q(p,t,t') \dots Q(p',t'',t''')  \rangle_c \approx \langle g(p,t,t') \dots g(p',t'',t''')  \rangle_c  
\eeq
\begin{displaymath}
\langle \delta \rho_k(t)\delta \rho_{k+p}(t')\dots \delta \rho_{k'}(t'')\delta \rho_{k'+p'}(t''')\rangle_c \approx 
\end{displaymath}
\beq
\approx \langle g(p,t,t') \dots g(p',t'',t''')  \rangle_c  \, H_k \dots H_{k'} \ .
\eeq
The above relationships hold at large distances, {\it i.e.} for small values of $p$, while the typical $k$'s are those of the static structure factor. A description in terms of an effective theory is valid provided the order parameter is small and this explains why the present treatment focus on the time scales of the $\beta$ regime.
I recall that in order to have small deviations  of the order parameter one needs to consider large enough wavelengths corresponding to $p$ small, indeed deviations on large values of $p$ are local and therefore large.  Once short-distance fluctuations are integrated out, Landau theory claims that the effective theory for the long-wavelength fluctuation takes a simple polynomial form.  For instance in the Curie-Weiss model the local spins are Ising and thus have a strongly non-linear distribution while the total magnetization becomes Gaussian in the thermodynamic limit.

In the field-theoretical description of dynamics of the previous section the order parameter can be written as $Q(ab)$ (I will first remove the $p$-dependence by considering the zero-dimensional case, space-dependence will be reintroduced later). Let us start considering the effective action in the FM limit. As we saw before, in this limit all dynamical correlations must assume a RS-like structure, therefore {\it the effective action in the FM limit must be also RS-like}. 
Furthermore the physics of the problem implies that the correlator at equal times is not  critical. In the context of supercooled liquids this follows from the known fact that the static structure factor $S_k$ is regular as a function of the temperature while the ergodicity breaking parameter $F_k$ displays a square-root singularity. We can thus set the fluctuations of $Q(aa)$ to zero without loss of generality.
No other symmetries should be imposed and thus we identify the effective theory with the generic replica-symmetric theory with $Q(aa)=0$. The corresponding action will be written explicitly below, see expression (\ref{action}).

The critical properties of this theory in the replica formalism ({\it i.e.} when  $a,b=1,\dots,n$) have been studied in \cite{Franz11b}. Unexpectedly it turned out that at all orders in the loop expansion the critical properties of the theory are the same of a quadratic stochastic equation.
This result is somehow spoiled by the fact that the quadratic stochastic equation is ill-defined beyond perturbation theory.
In the following I will also derive again this result diagrammatically. Within the diagrammatic treatment one can see that the very same mapping to an ill-defined {\it static} stochastic equation holds for dynamics in the FM limit. Indeed in the FM limit all correlators are RS-like and the only difference between dynamics and replicas occur when we perform integrals over the indexes, but, since we have $\int  da=1$ in both cases, the result is exactly the same.
As discussed in \cite{Rizzo14} and \cite{Rizzo15c} in order to go beyond this problem and obtain a well-defined theory near the avoided singularity one must abandon the FM limit and consider dynamics on a {\it finite}, albeit large, timescale, {\it i.e.} the $\beta$ regime.

On a finite time-scale the action must contain terms depending on $1/\Gamma_0$ that break the RS-like structure of the dynamical theory and restore time ordering leading to a correlation $C(t)$ that is a non-trivial function of time. This can be checked explicitly in mean-field SG models \cite{Kurchan92,Parisi13}.
Thus it is highly non trivial that we can study dynamics in the $\beta$ regime using the very same RS-like effective action valid in the FM limit, {\it i.e.} without the $\Gamma_0$-dependent terms. Indeed it was shown in \cite{Parisi13} that {\it the mean-field dynamical equations in the FM limit admit also non-FM solutions describing the time evolution of the correlator in the $\beta$ regime}.
When written explicitly the equations becomes identical to G{\"o}tze's equation for the critical correlator thus supporting the assumption that the GCT is the Landau theory corresponding to MCT.
All these technical results will be discussed again in details in the following sections.

At this point it should be clear that the connection between MCT and SBR is established at the level of the MCT equation for the critical correlator, without any reference to the microscopic derivation of MCT. The absence of the $\Gamma_0$ terms leads to a spurious time-scale invariance of the theory that can be removed by matching arguments. This time-scale invariance is a well-known feature of the MCT critical correlator.

\subsection{The Action and the Order Parameter}
In this section I define the mathematical problem I will address in this paper. 
I consider an order parameter that is a symmetric two-index object $Q(ab)=Q(ba)$ such that 
\beq
Q(aa)=0
\eeq
and I want to compute a generic average of the order parameter of the form:
\bew
\beq
\langle Q(ab) \dots Q(cd) \rangle \equiv {1 \over Z} \int \left(\prod_{(ab)} dQ(ab)\right) Q(ab)\dots Q(cd) \exp[-{\mathcal L}]
\label{genave}
\eeq 
Where the zero-dimensional GCT is defined as:
\beqa
{\mathcal L} & = & {1\over 2}\left(- \tau \int (da\,db)Q(ab)+m_2\int (da\, db\, dc)
Q(ab)Q(ac)+m_3\int (da\,db\,dc\,dd)Q(ab)Q(cd) \right)
\nonumber
\\
 & - &{1 \over
  6}w_1 \int(da\,db\,dc)Q(ab)Q(bc)Q(ca)-{1 \over 6}w_2
\int (da\,db)Q^3(ab) \ .
\label{action}
\eeqa
\eew
The coupling constants obey various constraints
\footnote{The expression of the bare propagator, to be presented below, implies $m_2 \leq 0$ and $m_2+m_3 \leq 0$. Without loss of generality I assume $w_1>0$.  Additional conditions also derived below imply $1/2<w_2/w_1<1$ that leads to $w_2>0$ and $w_1-w_2>0$.}.
The normalization constant $Z$ guarantees that $\langle 1 \rangle=1$.
Formally the same action describes the static and dynamical problem, the main difference being in the nature of the order parameter $Q(ab)$. An important difference is that in the dynamical case we need an {\it additional prescription} in order to make the problem of computing averages (\ref{genave}) with action (\ref{action}) well-defined. I will discuss this point later in the context of the mean-field equation.
I note that the symmetries of the action allow the presence of other types of cubic terms \cite{Temesvari02} but they are irrelevant for the final result as will be briefly explained in the appendix.

Let us discuss the differences between statics and dynamics. In the static treatment \cite{Franz11b,Szamel10} the indexes label different replicas $a=1,\dots,n$, and the limit $n \rightarrow 1$ must be taken at the end of the computation. As a consequence the order parameter is a $n \times n$ symmetric real-valued matrix.
In the dynamical formalism the indexes $a$ and the integral $\int da$ have the structure discussed previously and the order parameter $q(ab)$ is parameterized by four functions of the time coordinates $t_a$ and $t_b$. In both cases, however, we expect the {\it averages} of the order parameter to be {\it much simpler} objects.

In the static case the average order parameter is naturally a RS matrix described by a single scalar $q(ab)=q$,  $q(aa)=0$.
In the dynamical case we expect on physical grounds that the average order parameter obeys TTI and FDT and therefore it is parameterized by the form (\ref{2time}) with a function $C(t)$ such that $C(0)=0$. 
In the static case it is clear that the action is invariant under permutations of the replica indexes and the RS solution is an invariant too.  
In the dynamical case by analogy I will {\it define} a matrix $Q(ab)$ with the structure (\ref{2time}) an invariant. In the next section I will show that the invariant nature of the average order parameter can be proven perturbatively: first one shows that at the mean-field level the average can be described by an invariant, then one shows that corrections to the mean-field expression preserve the invariant character at all orders. It is likely that a non perturbative proof can be found but I leave this for future work \footnote{We note that in a full-fledged super-field formulation \cite{Zinn02,Kurchan92} one can derive FDT and TTI from symmetries of the action but here we choose the more compact setting of the first subsection, {\it i.e.} without fermions.}. 
In the following I will also use a notion of generalized invariant that applies to objects that depend on any  number of indexes $(a,\dots,c)$. In the replica case it is just a RS tensor while in the dynamical case the extension is slightly more complex and it is defined in the appendix.

A property of any invariant matrix $q(ab)$ with $q(aa)=0$ that will be {\it crucial} in the following developments is the following:
\beq
\int da\, q(ab)=0\ \ .
\label{zerocon}
\eeq
In the replica treatment the invariant is RS ($q(ab)=q$) and the result is just $(n-1)q$ that vanishes for $n=1$. 
In dynamics the physical origin of the condition can be traced back to FDT and TTI, at any rate it can be proven starting from the structure (\ref{2time}) and of $C(0)=0$ and it is discussed in the appendix \footnote{In the general case of invariant matrices with $q(aa)\neq 0$, $\int da q(ab)$ is equal to $q_d$ (the diagonal value of the matrix in the replica treatment) or to $C(0)$ in dynamics}. 
There are many properties ( {\it e.g.} $\int da=1$) like (\ref{zerocon})  that hold both in the replicated and dynamical formalism (also beyond FM) and allow a parallel treatment of the two problems. At the technical level this is one of the keys to the solution of the problem.

\section{The Loop Expansion}
\label{sloop}

In this section I will discuss the computation of the averages (\ref{genave}) by means of a loop expansion in order to compute systematically corrections to the mean-field result. I will first study the problem at the mean-field level, then study the bare propagator and finally setup the diagrammatic loop expansion. I will then show that the mapping to SBR holds at all orders for the most diverging terms.  In the end I will generalize the results to finite dimension adding a gradient term to the GCT. I will discuss at the same time the replica and dynamical case. In dynamics the action will be expanded around the non-FM solution of the mean-field equation of state corresponding to the equation for the critical correlator of MCT. These solutions do not have a RS-like form and  therefore the equivalence with the replica computation of \cite{Franz11b} will be lost.  

\subsection{The Mean-Field Solution}
\label{mfsol}

The mean-field value $q(ab)$ of the average order parameter is given by the solution of the equation $\partial {\mathcal L}/\partial q=0$:
\beqa
0 & = &-\tau + m_2 \int dc [q(ac)+q(bc)]+2 \, m_3 \int (dc\, dd) q(cd)+
\nonumber
\\
& - & w_1 (q^2)(ab)-w_2 \,q^2(ab)
\label{MFequaZD}
\eeqa
The above equation is valid for $a \neq b$, and in the summation over the indexes we must recall that $q(aa)=0$ \footnote{Note the presence of the factor $2$ in front of the $m_3$ term that is absent in the definitions of \cite{Rizzo13} and \cite{Parisi13}}.
The equation admits {\it invariant} solutions ({\it i.e.} RS in replica formalism or with the structure (\ref{2time}) in dynamics) and we will restrict the discussion to them. 
As a consequence the terms depending on $m_2$ and $m_3$ will not contribute because of the property (\ref{zerocon}) .
In the replica treatment the above equation for a RS solution $q(ab)=q$ reduces to
\beq
0=-\tau + 2 m_2 (n-1)q+2 \, m_3 n(n-1)q -w_1 (n-2)q^2-w_2 q^2
\eeq
giving for $n=1$ and positive values of $\tau$ \footnote{The positive solution is the stable one according to the assumption $w_1>0$} 
\beq
q=\sqrt{\tau \over w_1-w_2} \ .
\label{qstat}
\eeq
Thus the static treatment at the mean-field level implies the existence of a glassy solution for $\tau>0$ characterized by a squared root singularity at the critical value $\tau=0$.
In the dynamic case an invariant $q(ab)$ is parameterized by a real-valued function of time $C(t)$ and the equation can be rewritten as an equation for $C(t)$. As shown in \cite{Parisi13} and explained in the appendix the equation reads:
\beq
 \tau=-w_2 \, C^2(t)+w_1{d \over dt}\int_0^t C(t-t')C(t')dt'
\label{din1}
\eeq
This is the well-known G{\"o}tze equation for the critical correlator in Mode-Coupling-Theory.
An important feature of this equation is that it is time-scale invariant, in the sense that if $C(t)$ is a solution, $C(s\, t)$ is also a solution for any real $s$. Thus strictly speaking the mathematical problem defined by eqs. (\ref{action}) and (\ref{genave}) is ill-defined dynamically: we have a family of solutions related by a rescaling of time and we need a prescription to choose one of them. 
I will consider the following prescription:
{\it The action (and thus the mean-field equation) contains additional hidden terms that affect the small-time behavior of the solution leading to the condition 
\beq
\lim_{t \rightarrow 0} C(t) t^a=1
\label{prescr}
\eeq
}
where the value one is conventional (it can be replaced by any constant through a rescaling of 
time), but it is crucial that it does not depend on $\tau$. The exponent $a$ will be defined below. The prescription can be justified physically  with reference to the previous section. Indeed the hidden terms are those proportional to $1/\Gamma_0$ that appear when we abandon the fast motion limit. Their precise form is important for the dynamics on microscopic times. Instead, on the larger time-scale of the $\beta$ regime, their effect is just to remove the time-scale invariance and they can be replaced by the prescription (\ref{prescr}).
Furthermore, since the precise form of the $1/\Gamma_0$ terms is irrelevant, it is also clear that the precise nature of microscopic dynamics (Langevin vs. Hamiltonian) is irrelevant as well.

Le us recall the main properties of the solution of equation (\ref{din1}).
Depending on the sign of $\tau$ we have:
\beq
C(t)=\sqrt{|\tau|/w_1}\,g_{\pm}(t/\tau_\beta)\ ,
\label{scalfor}
\eeq
where the undetermined constant $\tau_\beta$ encodes the time-scale invariance of the solution.
The two functions $g_{\pm}(t)$ obey the following equation
\beq
\pm 1=-\lambda \, g_{\pm}^2(t)+{d \over dt}\int_0^t g_{\pm}(t-t')g_{\pm}(t')dt'
\label{ding}
\eeq
were the so-called parameter exponent is given by $\lambda=w_2/w_1$ \cite{Caltagirone12,Parisi13}.
The functions $g_{\pm}(t)$ diverges as $1/t^a$ for $t \rightarrow 0$. In the large time limit they behave differently, $g_{+}(t)$ goes to the constant value $1/\sqrt{1-\lambda}$ while $g_{-}(t)$ goes to minus infinity as $-t^{b}$. The exponents $a$ and $b$  are determined by the well-known expression 
\beq
\lambda={\Gamma^2(1-a)\over\Gamma(1-2a)}={\Gamma^2(1+b)\over\Gamma(1+2b)}\ .
\label{lambdaMCT}
\eeq
I recall that the above relationship implies additional constraints on $w_2/w_1$ \footnote{In particular $\lambda \leq 1$ (the equality corresponding to a tricritical point not described by the present theory) and $\lambda>1/2$ that follows from $b<1$ under the assumption of complete monotonicity (that is satisfied {\it e.g.} by correlation functions with Langevin Dynamics).}.
Note that the divergence of the correlator $g_{\pm}(t)$ at small times can be considered again a spurious consequence of time-scale invariance and, indeed, it is also fixed by the presence of the $1/\Gamma_0$ terms.

The above prescription in order to choose a unique $C(t)$ allows to express $\tau_\beta$ as a function of $\tau$ leading to the well-known MCT scaling law:
\beq
\tau_\beta =t_0 {1 \over \tau^{1 \over 2\, a}} 
\label{tauscal}
\eeq
where the constant $t_0$ is model-dependent.
I note that in both the static and dynamical treatment the deviations of the correlator from the plateau value are $O(|\tau|^{1/2})$ in the $\beta$ regime.
Instead,  we expect on physical grounds that at any {\it finite} time the variations of the correlator are linear with $\tau$, {\it i.e.} they vanish at leading order. This is the physical origin of the condition $q(aa)=0$: as I said already it corresponds to the well-known feature that the static structure factor in MCT is linear in $\tau$ at the transition. In the static treatment time differences are either zero $(a=b)$ or infinity ($a\neq b$) and $q(aa)=0$  is the only consequence. Instead,  in the dynamical treatment  time-scales are not clearly separated. This leads to the additional implication that the correlator $C(t)$ cannot be $O(|\tau|^{1/2})$ for $t\rightarrow 0$, and this implies that the prescription (\ref{prescr}) does not depend on $\tau$ leading to the scaling (\ref{tauscal}).

\subsection{The Bare Propagator}
\label{thebare}

In this subsection I will study the equation for the bare propagator of the theory. Here the time-scale invariance of the mean-field equation leads to the a spurious zero mode in the propagator that is removed by the prescription (\ref{prescr}). From this point on all results are presented for the first time.

By expanding the action (\ref{action}) around the solution of the mean-field equation (\ref{MFequaZD}) I obtain an equation for the bare propagator $G(ab)(cd)$:

\bew\beq
\int d(a' b')G(ab)(a'b') M_1(a'b')(cd)+m_2 \int dc'\left[ G(ab)(c'd)+ G(ab)(cc') \right]+2 m_3 \int(dc'dd')G(ab)(c'd')=\delta (ab)(cd)
\label{equaG}
\eeq
where $M_1(ab)(cd)$ comes from the quadratic part of the action, {\it i.e.}, it satisfies for a generic matrix $v(ab)$
\beq
\int d(cd) M_1(ab)(cd)v(cd) \equiv -2 w_2  q(ab)v(ab)- w_1 \int dc( q(ac)v(cb)+ v(ac)q(cb))
\eeq
\eew
The operators $M_1$ and $G$ are defined on the space of symmetric matrices with zero diagonal, therefore eq. (\ref{equaG}) and the above definition of $M_1$  hold only for $a\neq b$.
Equation (\ref{equaG})  must be supplemented with the condition 
\beq
G(ab)(cc)=0
\label{Gabcc}
\eeq
that follows from the constraint $Q(aa)=0$.
In the r.h.s. $\delta(ab)(cd)$ is a delta function in the space of symmetric matrices $Q(ab)$ with $Q(aa)=0$ such that $\int d(a'b') A(ab)(a'b')\delta(a'b')(cd)=A(ab)(cd)$ where the integration over the couples of different indexes can be written in terms of integrations over single indexes as  
\beq
\int d(ab)\dots \equiv {1 \over 2}\int da\,db\,(1-\delta(ab)) \dots
\label{restricted}
\eeq
Note the following relationship that will be essential in the following:
\beq
\int d(ab)=0
\label{sumzero}
\eeq
I recall that in the replica treatment the above result only holds for $n=1$, its general expression being $n(n-1)/2$. More technical details on the use of restricted ($d(ab)$) and unrestricted ($da \, db$) integrations are given later.

The exact expression for the bare propagator will be given at the end of this subsection, but it is instructive to consider the following quantity first:
\beq
\chi(ab) \equiv \int d(cd) G(ab)(cd)
\eeq
It is easy to see that $\chi(ab)$ is the (mean-field) longitudinal susceptibility, {\it i.e.} the derivative of the solution of the mean-field equation with respect to $\tau$:
\beq
\chi(ab)={d q(ab) \over d \tau} \ ,
\label{chi0}
\eeq
thus its expression can be obtained by derivation with respect to $\tau$ of the solution. 
In the replica case we have $\chi(ab)=\chi$ and from (\ref{qstat}) the derivative yields:
\beq
\chi={1 \over 2 \sqrt{\tau}}\left({1 \over w_1-w_2}\right)^{1/2}
\eeq 
In the dynamical case deriving expression (\ref{scalfor}) we obtain that $\chi(ab)$ is an invariant matrix parameterized by a function $\chi(t)$ given by:
\beq
\chi(t)={d C(t) \over d\tau}=\pm {1 \over 2 \sqrt{|\tau| w_1}}f_{\pm}(t/\tau_\beta)
\eeq
where the scaling function $f_{\pm}(x)$ is given by:
\beq
f_{\pm}(t)\equiv g_{\pm}(t)+{t \over a}\, \,{d{g}_{\pm}\over dt}(t) \ .
\label{chiscal}
\eeq
 the above scaling function has already been discussed in Ref. \cite{Berthier07}, (the additional factor $1/2$ in eq. (15) of \cite{Berthier07} is a typo). 
Since the leading correction to the small time behavior $1/t^a$ of $g_{\pm}(t)$ is $O(t^a)$ \cite{Gotze09} one obtains that $f_{\pm}(t)=O(t^a)$ at small times. For large times we have $f_-(t) \propto t^b$ and $f_+(t)=1/\sqrt{1-\lambda}$ (as in the static case). 

It is instructive to check that, as it should, the above expressions satisfy the equation for $\chi(ab)$ that can be obtained integrating over $d(ab)$ equation (\ref{equaG}). We have:
\bew\beq
\int d(a'b')\chi(a'b') M_1(a'b')(cd)+m_2 \int dc'\left[ \chi(c'd)+ \chi(cc') \right]+2 m_3 \int (dc'\,dd')\chi(c'd')=1\ .
\label{eqX}
\eeq\eew
Note that in both the replicas and the dynamical treatment we have 
\beq
\int db\,\chi(ab)=0 
\eeq
this is because in both cases $\chi(ab)$, being the derivative of an invariant matrix with zero diagonal, is itself an invariant matrix with zero diagonal and thus satisfies expression (\ref{zerocon}).
As a consequence, the two terms depending on $m_2$ and $m_3$ vanish.
In the replica case then it is immediate to check that the equation is verified while the dynamical case is less trivial. We can rewrite $\chi(ab)$ in the dynamical case as:
\beq
\chi(ab)={1 \over 2\, \tau} q(ab)+{1 \over 2 \, \tau\, a}q_0(ab)
\eeq
where $q_0(ab)$ is defined as the invariant matrix parameterized by $t\, dC(t)/dt$.
The first term is already a solution of  eq. (\ref{eqX}) because $q(ab)$ is a solution of the mean-field equation (\ref{MFequaZD}), thus:
\beq
\int d(cd) M_1(ab)(cd)q(cd)=-2 \, w_1 \, (q^2)(ab)-2 \, w_2\, q^2(ab)=2 \tau \ .
\eeq
It follows that $q_0(ab)$ must be, and indeed is, a {\it zero mode} of the operator $M_1$.
The reason why there is a zero mode lies in the time-scale invariance of the mean-field equation discussed previously. This can be seen starting from the fact that $C(s \,t)$ is a solution of the mean-field equation for any $s$ and deriving the mean-field equation with respect to $s$ at $s=1$. Therefore the equation  $\int d(a'b')\chi(a'b') M_1(a'b')(cd)=1$
leaves the constant in front of the zero mode unspecified.
In order to fix it we must resort to the prescription (\ref{prescr}) of the previous subsection. The very same $1/\Gamma_0$-dependent terms that control the small-time behavior of the mean-field equation will also be present in the equation for the bare propagator. Their effect will be to impose that $\chi(t)$ goes to zero at small times fixing the constant of the zero mode to $1/(2 \tau a)$. Consistently one may notice that when we obtained $\chi(ab)$ by derivation with respect to $\tau$ the second term in (\ref{chiscal}) came from the derivative of the time-scale $\tau_\beta$ which was previously fixed removing time-scale invariance using the small-time prescription. 

In order to proceed with the computation of the bare propagator let us observe that the mass matrix in the critical region (where $|\tau|$ is small) is made of a small $O(\sqrt{|\tau|})$ term ({\it i.e.} $M_1$) and two other terms that do not vanish as $\tau\rightarrow 0$. 
One should identify the modes to which these finite masses $m_2$ and $m_3$ correspond and argue that their fluctuations remain finite at $\tau=0$, {\it i.e.} they are not critical. From a critical point of view one would argue that all components of the bare propagator diverge as $O(1/\sqrt{|\tau|})$ except those corresponding to the massive modes that remain finite and can be considered to be zero on the scale $O(1/\sqrt{|\tau|})$ of the propagator.
As we will see in both statics and dynamics, what happens is quite different from these expectations and leads to the mapping to the stochastic equations.

A key observation is that the structure of the finite terms is the same in the replicated and dynamical formalism and we can build on results from the SG literature \cite{Temesvari02}. In particular we can rewrite the mass matrix $M(ab)(cd)$ as:
\beq
M(ab)(cd)=M_1(ab)(cd)- m_2  {\mathcal P}_A +4(m_2+m_3) {\mathcal N}_L
\eeq
where the operators ${\mathcal P}_A$ and ${\mathcal N}_L$ are defined (in both static and dynamics) as:
\beq
({\mathcal P}_A \phi)(ab) \equiv -\int dc\,(\phi(ac)+\phi(bc))+2 \int (dc \,dd)\phi(cd)
\eeq
\beq
({\mathcal N}_L \phi)(ab) \equiv \int d(cd)\phi(cd)
\eeq
One can check that ${\mathcal P}_A$ is a projector ${\mathcal P}_A^2={\mathcal P}_A$ on the so-called anomalous subspace (in the replica case this is true if $n \rightarrow 1$, in dynamics it follows from the condition $\int da=1$).
This is the space of matrices $Q(ab)$ with $Q(aa)=0$ that can be written (for $a\neq b$) as $Q(ab)=\psi(a)+\psi(b)$ with $\int da \psi(a)=0$.
The operator ${\mathcal N}_L $ in matrix form is a constant {\it i.e.} ${\mathcal N}_L (ab)(cd)=1$: it transforms any matrix into a constant but it is {\it not} a projector, instead it is nilpotent ${\mathcal N}_L^2=0$ . In both dynamics and statics this property follows from eq. (\ref{sumzero}) and it is essential in the following developments, we will see that the ${\mathcal N}_L $ term will increase fluctuations instead of reducing them.

To proceed let us make some qualitative considerations. Given that the anomalous sector has a finite mass we would expect that the leading divergent behavior of the propagator would not be changed if we just put the corresponding component to zero. More precisely we could write:
\beq
{\mathcal P}_A G=O(1)
\eeq
that is equivalent to:
\beq
\int dc\  G(ab)(cd)+O(1)=\int (dc\, dd)G(ab)(cd)\equiv 2 \chi(ab)\ .
\eeq
If we substitute the above expressions into eq. (\ref{equaG}) we have:
\beq
\int d(a'b')G(ab)(a'b') M_1(a'b')(cd)+4 (m_2 + m_3) \chi(ab) = O(1)
\eeq
now multiplying by $M_1^{-1}$ we have:
\bew
\beq
G(ab)(cd)= -4 (m_2 + m_3) \chi(ab) \int d(a'b') M_1^{-1}(a'b')(cd) +O(M_1^{-1}) \ .
\eeq
\eew
Multiplying by $M_1^{-1}$ the equation $\int d(ab)\chi(ab) M_1(ab)(cd)=1$ we have 
\beq
\int d(a'b') M_1^{-1}(a'b')(cd)=\chi(cd)
\eeq
from which we finally arrive at:
\beq
G(ab)(cd)= -4 (m_2 + m_3) \chi(ab)  \chi(cd)+ O(M_1^{-1})
\eeq
this implies that $G(ab)(cd)$ has a double pole {\it i.e.}, while the small  mass $M_1$ is $O(\sqrt{\tau})$ and thus the longitudinal susceptibility $\chi(ab)$ is $O(M_1^{-1})= O(1/\sqrt{|\tau|})$, the full propagator is $O(1/|\tau|)$, {\it i.e.} it is proportional to the squared inverse mass.
This has to be contrasted with the following integrals of the propagator
\beqa
\int (dc\, dd)G(ab)(cd) & \equiv &  2\,\chi(ab)
\\
\int\,dc\,G(ab)(cd) & \equiv &  2\,\chi(ab)+O(1)
\eeqa
that are less divergent, note that upon integration the double-pole contribution in $G(ab)(cd)$ vanishes because of $\int da \chi(ab)=0$ according to (\ref{zerocon}).

We will now present the exact expression of the bare propagator. We define:
\beq
\Delta M \equiv  M_1(ab)(cd)- m_2 {\mathcal P}_A
\label{DELTAM}
\eeq
therefore the equation for the propagator becomes:
\beq
(\Delta M - \,  v\,  {\mathcal N}_L)G=I  
\label{eqGcomp}
\eeq
with the following definion:
\beq
v \equiv -4(m_2+m_3)\ . 
\label{defv}
\eeq
The {\it exact} solution is
\beq
G(ab)(cd)={1 \over \Delta M}(ab)(cd)+ v \, \chi(ab) \chi(cd)
\label{propa}
\eeq
where $\chi$ is the longitudinal susceptibility defined before, {\it i.e.} it is the solution of the following equation  in the space of symmetric matrices zero on the diagonal
\beq
\chi \cdot \Delta M= \chi M_1=n_L
\eeq
where $n_L$ is defined as the matrix with all elements equal to one $n_L(ab)=1$, $n_L(aa)=0$.
Equation (\ref{eqGcomp}) holds because we have ${\mathcal N}_L \chi={\mathcal P}_A \chi=0$, $ ({\mathcal N}_L \Delta M^{-1})(ab)(cd)= n_L(ab) \chi(cd)$ which cancels with the term $(\Delta M \chi)(ab) \chi(cd)= n_L(ab) \chi(cd)$.

Note that we have derived above the explicit expression for $\chi(ab)$ but
we {\it cannot} write down in explicit form neither $1/\Delta M$ nor $1/M_1$. However, in order to proceed we do not need these quantities explicitly.
{\it The only information we will use is that $1/\Delta M$ is $O(1/\sqrt{|\tau|})$ for $|\tau|$ small, {\it i.e.}, it is less divergent than the leading order $O(1/|\tau|)$ term}.
This follows because: i) $1/M_1$ scales {\it exactly} as $O(1/\sqrt{|\tau|})$ because the mean-field solution $q(ab)$ is proportional to $\sqrt{|\tau|}$, ii) ${\mathcal P}_A$ is a projector and thus it can reduce fluctuations in the anomalous sector without changing the scaling of the components that remain critical. 

Summarizing, the key point is that ${\mathcal P}_A$ is a projector, whereas the operator ${\mathcal N}_L$ is not. As a consequence, while the presence of  ${\mathcal P}_A$ does not change the scaling of $1/\Delta M$ with respect to $1/M_1$, the presence of ${\mathcal N}_L$ leads to a completely different scaling behavior of $G$ with respect to $1/\Delta M$ with appearance of a double pole.

\subsection{The Diagrammatic Loop Expansion}
\label{diagram}

Let us rewrite the Glassy Critical Theory in terms of the deviation of the order parameter $Q(ab)$ from the mean-field solution $q(ab)$ obtained before. The action (\ref{action}) reads:
\bew
\beq
{\mathcal L}= {1 \over 2} (\delta Q G^{-1} \delta Q)- {1 \over
  6}w_1 \int(da\,db\,dc)\delta Q(ab) \delta Q(bc)\delta Q(ca)-{1 \over 6}w_2
\int (da\,db)\delta Q^3(ab) \ .
\label{acexp}
\eeq 
\eew
where $\delta Q(ab)\equiv Q(ab)-q(ab)$ and $G(ab)(cd)$ is the bare propagator studied above.
We consider now the loop expansion obtained by using the Taylor series for the terms proportional to $w_1$ and $w_2$. The various terms in the expansion are given by all Feynman diagrams with vertices of degree three.
The diagrams contributing to a correlation function of the form (\ref{genave}) involving the order parameter to the power $E$ are given by the (infinite) set of diagrams with $E$ external lines. 
The generic average of the form (\ref{genave}) picks up contributions from both connected and disconnected diagrams. 
Diagrams where there is a disconnected component that does not contain any external line are called 
fully disconnected diagrams and do not contribute, being canceled by the normalization factor (see {\it e.g.} \cite{Parisi88}).
\begin{figure}
\includegraphics[width=10cm]{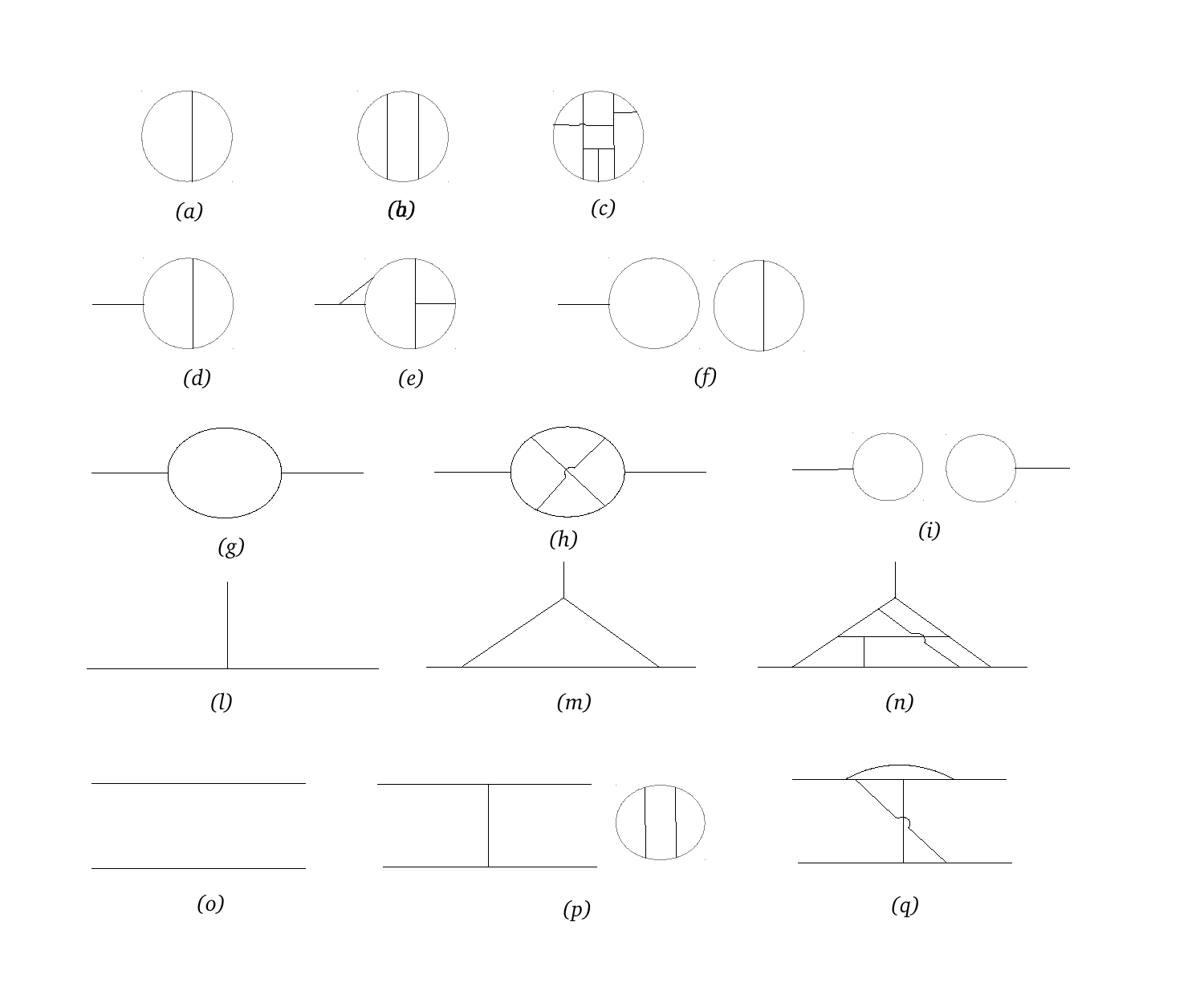}
\caption{Some of the (infinite) cubic diagrams for the computation of $Z$ and to correlations with the order parameter to the power $E=1,2,3,4$. Diagrams $(i)$ and $(o)$ are disconnected and contribute to the correlations but not to connected correlations. Diagrams $(f)$ and $(p)$ are fully disconnected and does not contribute at all. }
\label{f1}
\end{figure}
Connected correlations obtained from the generic correlation (\ref{genave}) corresponds to connected diagrams. 

In fig. (\ref{f1}) we display some of the Feynman diagrams that need to be considered for various objects.  To any line in a given diagram we must attach the bare propagator (\ref{propa}) and to any vertex we must attach the contributions given by $w_1$ and $w_2$ in (\ref{acexp}). 
For instance if we consider the diagram $(g)$  of fig. (\ref{f1}) that contributes to the dressed propagator $\langle Q(ab) Q(cd)\rangle_c$ we will have four contributions: 
\bew
\beqa
& & w_1^2\, \int (dm\,dn\,dl) (dm'\,dn'\,dl')\, G(ab)(mn)\, G(lm)(l'm')  \, G(ln)(l'n') \, G(m'n')(cd)   +     
\nonumber
\\  
& + & w_1\,w_2 \,  \int (dm\,dn\,dl) (dm'\,dn')\,G(ab)(mn)\, G(lm)(m'n')  \, G(ln)(m'n') \, G(m'n')(cd)   +       
\nonumber
\\  
& + & w_1\,w_2 \, \int (dm'\,dn'\,dl') (dm\,dn)\,G(ab)(mn)\, G(mn)(l'n')  \, G(mn)(m'l') \, G(m'n')(cd)   +      
\nonumber
\\  
& + & w_2^2 \, \int (dm'\,dn') (dm\,dn)\,G(ab)(mn)\, G(mn)(m'n')  \, G(mn)(m'n') \, G(m'n')(cd)              \ .
\label{digex}
\eeqa
\eew
Note that the action (\ref{action}) is written in terms of {\it unrestricted} integrals over the coordinates $a,b,c$, {\it i.e.} in principle it depends also on $Q(aa)$, therefore the condition $Q(aa)=0$  must be explicitly enforced whenever we use unrestricted integrations as we did below eq. (\ref{MFequaZD}) and in (\ref{Gabcc}). When evaluating the Feynman diagrams we can use the unrestricted expressions for the vertexes (as we did above in (\ref{digex})) because the constraint $\delta Q(aa)=0$ is automatically enforced by the propagator through the condition $G(ab)(cc)=0$. 

Given that the propagators attached to the lines diverge as $1/|\tau|$ we expect that each diagram will also be divergent. In particular the naive expectation would be that each of the four terms  in expression (\ref{digex})  should  be divergent as $|\tau|^{-4}$. Instead, this is wrong, because it turns out that the leading $O(\tau|^{-4})$ term has a zero prefactor.
In order to see this let us replace $G(ab)(cd)$ in the above expressions with its most diverging term $v \chi(ab)\chi(cd)$ according to (\ref{propa}). We can check that the summation at each vertex can now be done independently. For a vertex of type $w_1$ we have a contribution given by:
\beq
\int (dm\,dn\,dl) \chi(mn)\chi(ml)\chi(ln)\ ,
\label{prew1}
\eeq
while for a vertex of type $w_2$ we have a contribution 
\beq
\int (dm\,dn) \chi^3(mn)\ ,
\label{prew2}
\eeq
 the key point is that both contributions vanish. 
This can be seen starting from the property (proved in the appendix) that given two invariant matrices $A(ab)$ and $B(ab)$ both $A(ab)B(ab)$ and the matrix product $(AB)(ab)$ are also invariant matrices. 
Since $\chi(ab)$ is a zero-diagonal invariant matrix, $\chi^3(ab)$ is also a zero-diagonal invariant matrix, therefore the property (\ref{zerocon}) implies that the prefactor (\ref{prew2}) vanishes.
The prefactor (\ref{prew1}) can be rewritten in terms of the matrix $ \chi(mn)(\chi^2)(mn)$ which is zero on the diagonal  and property (\ref{zerocon}) can again be used. We have thus shown that 
\beqa
\int (dm\,dn\,dl) \chi(mn)\chi(ml)\chi(ln) & = & 0
\\
\int (dm\,dn) \chi(mn)(\chi^2)(mn) & = & 0
\eeqa

\begin{figure}
\includegraphics[width=9cm]{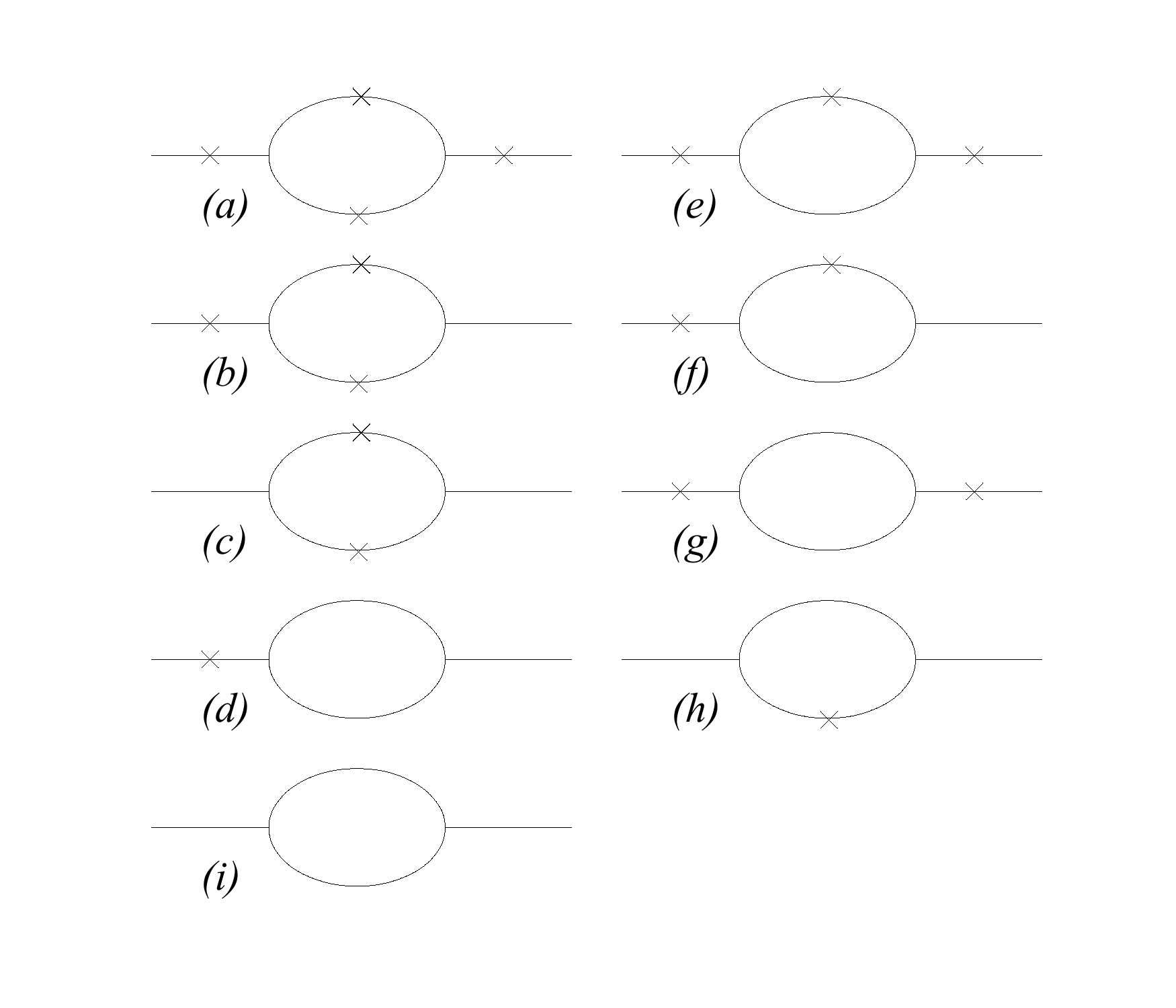}
\caption{Diagrams contributing to the two point correlation in the crossed lines representation.
Seven other diagrams can be obtained from symmetries.
The necessary  condition implies that diagrams $(a)$ and $(b)$ vanish, the necessary and sufficient condition implies that diagrams $(g)$ and $(e)$ also vanish. All remaining diagrams give diverging contributions with non-zero prefactors. The maximal degree of divergence is given by the terms where the full lines are organized on trees, {\it i.e.} diagrams $(f)$ and $(c)$.}
\label{f}
\end{figure}

In order to see which contributions are non-zero let us now introduce a graphical representation of the propagator $G(ab)(cd)$
by associating the most divergent term $v \chi(ab)\chi(cd)=O(|\tau|^{-1})$ to a crossed line and the less divergent term $\Delta M^{-1}=O(|\tau|^{-1/2})$ to a full line. This representation is borrowed from the literature on the Random-Field Ising Model (RFIM), see {\it e.g.} \cite{DeDominicis10}.
Now the diagram, say, $(g)$ of fig. (\ref{f1}) can be rewritten as the sum of 16 diagrams (see fig. (\ref{f}) where only nine diagrams are shown because the remaining seven can be obtained by symmetries).
The most diverging contribution is obtained by putting crossed lines on all lines, this corresponds to diagram $(a)$ in fig. (\ref{f}).  However, we have already seen that if we put a crossed line on each of the three legs of a given vertex we will have a vanishing contribution, either  (\ref{prew1}) or (\ref{prew2}). The above property leads to the following {\it necessary} condition to have a non-zero contribution: each vertex must have at least a full line attached to it.
This implies that both diagrams $(a)$ and $(b)$ in fig. (\ref{f}) vanish.

In order to obtain the maximum degree of divergence of a given cubic diagram we have to maximize the number of crossed-lines while satisfying the above necessary condition.  The necessary condition can be fulfilled by connecting the vertices through tree-like (no-loop) structures of full lines. Indeed, suppose that there is a loop of full lines on the Feynman diagram (as in diagrams $(g)$, $(d)$ and $(i)$ in fig. (\ref{f})), we can open the loop by replacing one full line with a crossed line obtaining a more divergent diagram that still verifies the necessary condition that each vertex has a full line attached to it (respectively $(f)$, $(e)$ and $(h)$).

Actually, the condition that each vertex must have at least a full line attached to it is necessary but not sufficient to give a non-zero contribution. The necessary and sufficient condition, that will be discussed in the appendix, is that if we split the graph into subgraphs by cutting all crossed lines, every subgraph must contain at least one external line.
Therefore {\it the condition of maximal divergence is that each vertex in the diagram must be on a tree of full lines and each tree must be connected to an external line}.
With respect to figure (\ref{f}) we have the following: 
the necessary  condition implies that diagrams $(a)$ and $(b)$ vanish, the necessary and sufficient condition implies that diagrams $(g)$ and $(e)$ also vanish. All remaining diagrams give diverging contributions with non-zero prefactors. The maximal degree of divergence is given by the terms where the full lines are organized on trees, {\it i.e.} diagrams $(f)$ and $(c)$. 
It should be clear that the previous discussion applies to any cubic diagram with any number of external legs and any number of vertices, {\it i.e.} at all orders in the loop expansion.

\subsection{Mapping to the Stochastic equation}

In the previous subsection we have shown that the leading divergent diagrams are those associated to trees and now we will proceed to show that these diagrams are precisely those that are generated by the solutions of a stochastic equation.

Let us rewrite the quadratic part of the action (\ref{action}) in terms of the operators ${\mathcal P}_A$ and ${\mathcal N}_L$ introduced in subsection (\ref{thebare}) as:
\bew
\begin{displaymath}
 {1\over 2}\left(- \tau \int (da\,db)Q(ab)+m_2\int (da\, db\, dc)
Q(ab)Q(ac)+m_3\int (da\,db\,dc\,dd)Q(ab)Q(cd) \right)=
\end{displaymath}
\beq
 =  -\tau \int d(ab) Q(ab)-{m_2 \over 2} (Q {\mathcal P}_A Q) +{v \over 2} \left( \int d(ab)Q(ab) \right)^2
\eeq
The last term can be rewritten in terms of random shift of $\tau$ using the identity:
\beq
\exp \left[ -{v \over 2} \left( \int d(ab)Q(ab) \right)^2\right]=\int {ds \over \sqrt{2 \pi v}}e^{-{s^2\over 2 v}} \exp \left[ s\,\int d(ab)Q(ab)\right] 
\label{rewrite}
\eeq
and the generic correlation (\ref{genave}) can now be rewritten as
\beq
\langle Q(ab)\dots Q(cd)\rangle=\frac{\overline{\langle Q(ab)\dots Q(cd)\rangle_s Z_s}}{\overline{Z_s}}
\label{annave}
\eeq
where the overline means average with respect to a Gaussian variable $s$ with zero mean and variance $v$. The suffix $s$ in the averages $\langle \dots \rangle_s$ and in $Z_s$ means that they are evaluated with a weight $\exp[-{\mathcal L}_s]$ with:
\beqa
{\mathcal L}_s & \equiv & - (\tau+s) \int d(ab)Q(ab)-{m_2 \over 2} (Q {\mathcal P}_A Q)+
\nonumber
\\
 & - &{1 \over
  6}w_1 \int(da\,db\,dc)Q(ab)Q(bc)Q(ca)-{1 \over 6}w_2
\int (da\,db)Q^3(ab)
\eeqa
\eew
With the above transformation the analogy with a stochastic equation is much more plausible but one should also recognize that the two objects are essentially different, and indeed, as we will see, only their leading divergent behavior is the same.
An essential property is that independently of $s$ we have $Z_s=1$ (and also $Z=1$) {\it exactly} and therefore it can be dropped from the above expression giving:
\beq
\langle Q(ab)\dots Q(cd)\rangle=\overline{\langle Q(ab)\dots Q(cd)\rangle_s}
\label{corexa}
\eeq
This implies that while the mapping to a stochastic equation is approximate {\it there is instead an exact mapping to a problem with Gaussian quenched disorder}.
The condition $Z_s=1$ is essential in this mapping because otherwise the disorder $s$ would not be an independent Gaussian variable, {\it i.e.} it would not be quenched.
The proof of the identity $Z_s=1$  is postponed to the appendix. We note however that in the replicated case is almost trivial. It follows from $\ln Z=O(n-1)$ which is valid at all orders in the loop expansion. In the dynamical context it is closely related to the system being at equilibrium and does not hold for off-equilibrium dynamics.
A detailed discussion of the mapping to the quenched problem will be given in the next section. In the following we will use it to complete the perturbative analysis of the previous subsection.

\begin{figure}
\includegraphics[width=9cm]{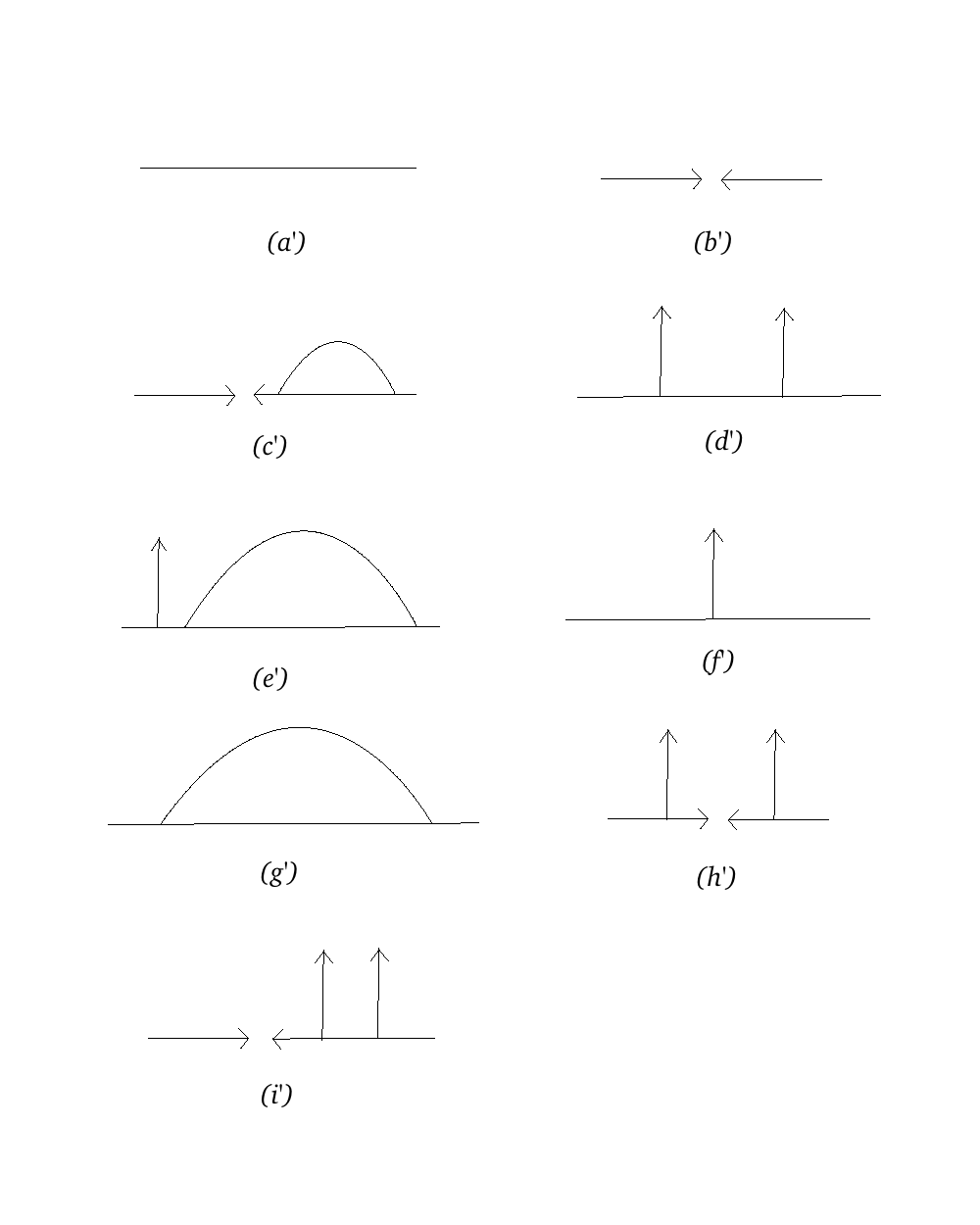}
\caption{Diagrams contributing to the loop expansion 
of $\langle Q(ab)Q(cd)\rangle_s$ before averaging.}
\label{preave}
\end{figure}

Let us consider the expression of the correlations $\langle Q(ab) \dots Q(cd) \rangle_s$ before averaging over the disorder.
They can be computed in a loop expansion as well. Although $s$ and $\tau$ play the same role in ${\mathcal L}_s$, in order to make contact with the loop expansion of the GCT it is appropriate to study the expansion of the correlations in powers of the source $s$ and of the coupling constants $w_1$ and $w_2$. 
The mean-field solution of the action ${\mathcal L}_0$ with zero source is therefore the same of ${\mathcal L}$. Indeed they differ by a term $ -{v \over 2} \left( \int d(ab)Q(ab) \right)^2$ that, as we saw before, is irrelevant for the solution of the equation. The difference becomes apparent at the level of the bare propagator which is now given  by the  sole term $1/\Delta M$. Therefore the correlation function of a product of $E$ $Q$'s  before averaging is given by the sum of all cubic diagrams with $E$ external lines,  bare propagators $1/\Delta M$, the same vertices  $w_1$ and $w_2$ and  sources $s$. The sources will be represented by arrows and we must attach to the body of the arrows a bare propagator $1/\Delta M$ and to the head the field $s$.
In figure (\ref{preave}) we display some of the diagrams contributing to the two-point correlation $\langle Q(ab)Q(cd)\rangle_s$. The diagram $(a')$ corresponds to the bare propagator $1/\Delta M$.
Note that the integration of the indexes at the head of the arrow can be done explicitly and we obtain for the total contribution of an arrow
\beq
\int d(cd) {1 \over \Delta M}(ab)(cd)\, s=s\, \chi(ab)
\eeq

We can now see how the diagrammatic expression of  $\langle Q(ab)\dots Q(cd)\rangle$  in terms of full and crossed lines obtained in the previous subsection can be also obtained from the above diagrammatic expression of $\langle Q(ab)\dots Q(cd)\rangle_s$ in terms of full lines and arrows {\it after averaging} over the random term $s$.
Indeed, the Gaussian average over $s$ selects the diagrams with even powers of $s$, each couples of sources gives a contribution $\overline{s^2}=v$, this can be represented graphically by saying that two arrows, each one with an associated term $s \chi$, are joined to form a crossed line associated to a term $v \chi \chi$.
Let us consider explicitly a few examples: i) diagrams $(a')$ and $(b')$ give precisely the bare propagator of the GCT, eq. (\ref{propa}), ii) diagrams $(e')$ and $(f')$ vanish because the average of $s$ is zero, iii) the other diagrams corresponds to diagrams of the GCT shown in fig. (\ref{f}): $(c')\rightarrow (d)$, $(d')\rightarrow (h)$, $(g')\rightarrow (i)$, $(h')\rightarrow (c)$, $(i')\rightarrow (b)$. 

Up to this point no approximation has been made, we have just explained how the {\it exact} equation (\ref{corexa}) can be verified at all orders comparing the diagrammatic expansions of its left-hand and right-hand sides. In the previous subsection, however, we have argued that in the critical region, where the bare propagator $G(ab)(cd)$ is large, the most diverging contributions to  a given cubic diagram (those of fig. (\ref{f1}))  are those where each vertex belong to tree-like networks of full lines $1/\Delta M$ in the crossed-lines representation (as in fig. \ref{f}).
Since the process of averaging over $s$ does not alter the structure of the full lines $1/\Delta M$ this implies that the most divergent diagrams had a tree-like structure also before the $s$ averaging, meaning that we can replace $\langle Q(ab)\dots Q(cd) \rangle_s$ with its expression computed only with tree-like diagrams.
\beq
\langle Q(ab)\dots Q(cd)\rangle \approx \overline{\langle Q(ab)\dots Q(cd)\rangle_s^{tree}}
\label{cortree}
\eeq
A classic field-theoretical result is that the sum of all tree-like diagrams is given exactly  by the following expression
\beq
\langle Q(ab)\dots Q(cd)\rangle_s^{tree}=q_s(ab) \dots q_s(cd)
\eeq
where $q_s(ab)$ is the solution of the mean-field equation {\it in presence} of the source $s$
\beq
0=-(\tau+s) - m_2 ({\mathcal P}_A q_s)(ab) -w_1 (q_s^2)(ab)-w_2 \,q_s^2(ab)
\label{stoch}
\eeq
In the dynamical case we must remember that the above equation is time-scale invariant and leaves $q_s(ab)$ undetermined up to a rescaling of time. As discussed before this invariance is removed by the prescription (\ref{prescr}) that follows from the presence of hidden terms in the action controlling the small-time behavior of the order parameter. Since ${\mathcal L}_s$ is obtained from ${\mathcal L}$ through (\ref{rewrite}), one can easily see that these terms are also present
in ${\mathcal L}_s$  and therefore the very same prescription (\ref{prescr}) must be used to determine $q_s(ab)$.

Let us write down the result explicitly in the static and dynamical case.
In the static case for a correlation of order $E$ we have
\beq
\langle Q(ab)\dots Q(cd)\rangle \approx \overline{q^E_s}
\label{repzero}
\eeq
 where $q_s$ is the solution of 
\beq
0=-(\tau+s)  +(w_1-w_2) \,q_s^2 \rightarrow q_s=\sqrt{\tau+s \over (w_1-w_2)}
\eeq
Thus we have rederived diagrammatically the result of \cite{Franz11b}.
Note that a real solution $q_s$ to the equation exists only for $\tau+s \geq 0$, therefore the expression  $\overline{q^E_s}$ is ill-defined because it involves an integration over $s$ on the whole real axes. This suggests that $\langle Q(ab)\dots Q(cd)\rangle$ is actually ill-defined in the replica case.
Instead, in the dynamical case if we specialize to the correlation component of the order parameter $Q(ab)$ (no $\eta$'s) we have:
\beq
\langle C(t_a,t_b)\dots C(t_c,t_d) \rangle \approx \overline{C_s(t_a-t_b)\dots C_s(t_c-t_d)}
\label{welldef}
\eeq
where $C_s(t)$ is solution of the time-scale invariant dynamical equation
\beq
 \tau+s=-w_2 \, C_s^2(t)+w_1{d \over dt}\int_0^t C_s(t-t')C(t')dt'
\label{dins}
\eeq
with the prescription (\ref{prescr}).
The function $C_s(t)$ exists for all values of $s$ and therefore the average (\ref{welldef}) is perfectly well defined, at variance with the replica expression (\ref{repzero}). The full solution can easily be written in terms of the two solutions $g_{\pm}(t)$ of (\ref{ding}):
\beq
C_s(t)=\sqrt{|\tau+s|/w_1}g_{\sign(\tau+s)}(t/(|\tau+s|/w_1)^{1/2 a})
\eeq
The above equations define SBR in the zero-dimensional case. We have thus proven that at each order in the loop expansion the leading divergent terms of the GCT are the same of SBR, as anticipated in \cite{Rizzo14}.

In the replica case a correlation $\langle Q(ab)\dots Q(cd)\rangle$ of order $E$ takes different values depending on the number of indexes that are different. Note however that  these differences do not show up at the leading order because they are all given by $\overline{q^E_s}$ according to (\ref{repzero}). These features are corrected by subleading terms that cannot be associated to the stochastic equation \footnote{Actually some less divergent correlations of order $E$ can be obtained from the stochastic equation by deriving $k$ times with respect to $\tau$ the correlations of order $E-k$. These kind of correlations are called {\it connected} in the RFIM literature }. A similar feature is present in the dynamical case: the l.h.s. of (\ref{welldef}) depends explicitly on all time differences while the r.h.s. depends only on $t_a-t_b, \dots, t_c-t_d$.  Again this property  does not hold exactly due to subleading terms that cannot be mapped to a stochastic equation.

\subsection{Finite Dimension}

We now add a space-dependence to the order parameter $Q_x(ab)$ with the same properties described in the previous sections. In Landau theory the standard way to modify the action is to add a term proportional to the squared gradient of the order parameter, arguing that all other terms depending on spatial derivatives are less important in the critical region. Thus we are now interested in computing the averages 
\bew
\beq
\langle Q_x(ab) \dots Q_y(cd) \rangle \equiv {1 \over Z_D} \int \left(\prod_{(ab),z} dQ_z(ab)\right) Q_x(ab)\dots Q_y(cd) \exp[-\int dx' {\mathcal L}_{x'}]
\label{genavespace}
\eeq 
\eew
where 
\beq
{\mathcal L}_{x} \equiv  {1 \over 2}  \int d(ab)|\nabla Q_x(ab)|^2+  {\mathcal L}[Q_x] 
\eeq
and $ {\mathcal L}[Q_x] $ means the action (\ref{action}) evaluated for $Q(ab) \rightarrow Q_x(ab)$.
In order to setup the perturbative computation of the above averages we perform the same steps of the previous section.

The mean-field equation is the same of (\ref{MFequaZD}) plus an additional Laplacian term $\nabla^2 Q_x(ab)$.
I will consider a translational invariant solution $Q_x(ab)=q(ab)$. Thus the Laplacian is irrelevant and the solution $q(ab)$ is the same of section (\ref{mfsol}).  Recall once again that the mean-field equation is time-scale invariant and the action must be supplemented with the prescritpion (\ref{prescr}) to have a unique mean-field solution. In the space-dependent context we have to generalize the prescription by assuming that the same local terms are present at each point in space in order to have the same prescription everywhere and thus a constant mean-field solution.

The bare propagator instead changes with respect to the zero-dimensional case. As susual it is conveniently diagonalized in Fourier space:
\beq
G_{k,k'}(ab)(cd)=G_k(ab)(cd)\delta(k+k')\ ,
\eeq
where $G_k$ satisfies the equation 
\beq
(k^2+\Delta M+v {\mathcal N}_L)G_k=I
\eeq 
that generalizes eq. (\ref{eqGcomp}). The quantities $\Delta M$, ${\mathcal N}_L$ and $v$ have the same definitions of eq. (\ref{eqGcomp}). The {\it exact} solution is 
\beq
G_k(ab)(cd)={1 \over k^2+ \Delta M}(ab)(cd)+ v \, \chi_k(ab) \chi_k(cd)
\eeq
where $\chi_k(ab)$ is the solution of the equation
\beq
\chi_k (k^2+ \Delta M)=1\ .
\label{chik}
\eeq
The physical meaning of the function $\chi_k$ can be understood considering the response of the system to a space-dependent perturbation on $\tau$. This can be implemented 
by  the following replacement in the expression of the action:
\beqa
 & & \tau \int dx \left[ \int d(ab) Q_x(ab)\right]  \rightarrow 
\int\, dx \, \tau_x\,\left[ \int d(ab) Q_x(ab)\right] 
\nonumber
\\
& & =\int dk \, \tau_{-k}\,\left[ \int d(ab) Q_k(ab)\right] \ .
\eeqa
A uniform value of $\tau_x$  corresponds in Fourier space to $\tau_k=\tau_0 \, \delta(k)$, meaning that all components of the mean-field solution but $k=0$ vanish. A non-constant $\tau_x$ induces a non-constant mean-field solution meaning that the components $q_k(ab)$ is different from zero also for $k \neq 0$.
One can see that the susceptibility of $q_k(ab)$ to $\tau_{k'}$ at the constant solution $\tau_x=\tau$ is determined by $\chi_k$, more precisely we have:
\beq
\left. {d q_k \over d \tau_{k'}} (ab) \right|_{\tau_x=\tau}=\delta(k+k')\chi_k(ab)
\eeq
The above relationship generalizes eq.  (\ref{chi0}) and can be verified by deriving the mean-field equation with respect to $\tau_{k'}$ obtaining thus eq.  (\ref{chik}) for the susceptibility.

Given that, both in the static and dynamical case, the solution $q_k(ab)$ in presence of a inhomogeneous $\tau_x$ is an invariant, the matrix $\chi_k(ab)$ is also an invariant.
In the dynamical treatment the corresponding function $\chi_k(t)$ obeys the following equation
\beq
k^2 \chi_k(t)-2 w_2 C(t)\chi_k(t)+2 w_1 {d \over dt}\int_0^t C(t-t')\chi_k(t')dt'=1\ .
\label{chikcor}
\eeq
At variance with the $k=0$ case the function $\chi_k(t)$ does not have a simple expression in terms of the mean-field solution $C(t)$. Note that $\chi_k(t)$ should be naturally identified with the scaling function for the so-called three-point susceptibility in the $\beta$ regime in the context of IMCT, although eq. (\ref{chikcor}) does not appear explicitly in \cite{Biroli06}. 
A detailed discussion of $\chi_k(t)$ is not essential here, indeed to proceed we only need the property that $\chi_k(ab)$ is an invariant.

We can now repeat the analysis of the leading divergences of the diagrams, the only difference being that  each vertex carries a momentum dependence to be integrated upon. Before the momentum integration we can already see that the contribution of a vertex that has three crossed lines at his legs is zero. Indeed we have the two contributions
\beq
w_2 \int d(ab) \chi_k(ab) \chi_{k'} (ab)  \chi_{-k-k'} (ab) \ ,
\eeq
\beq
w_1 \int (da\, db\, dc) \chi_k(ab) \chi_{k'} (bc)  \chi_{-k-k'} (ca) 
\eeq
and again both vanish because $\chi_k(ab)$ is an invariant with zero diagonal. Therefore the same necessary condition leads to the choice of tree-like structures in order to maximize the degree of divergence of a graph.
The analysis of the zero-dimensional case can now be extended in a straightforward fashion.

One rewrites the quadratic term in the GCT introducing a quenched space-dependent Gaussian fluctuation $s(x)$ as in (\ref{rewrite}) 
\beq
\overline{s(x)s(y)}=\delta (x-y)\, v \ .
\eeq
In this way the averages (\ref{genavespace}) can be rewritten as in (\ref{annave}) 
\beq
\langle Q_x(ab)\dots Q_y(cd)\rangle=\frac{\overline{\langle Q_x(ab)\dots Q_y(cd)\rangle_s Z_s}}{\overline{Z_s}}
\label{annavespace}
\eeq
Once again it turns out that the partition function is equal to one independently of the realization of the random fluctuations: $Z_s=1$. This leads to the fundamental result that {\it the GCT is exactly equivalent to a theory with quenched disorder}:
\beq
\langle Q_x(ab)\dots Q_y(cd)\rangle=\overline{\langle Q_x(ab)\dots Q_y(cd)\rangle_s}
\label{corexaspace}
\eeq   
The perturbative loop expansion of the GCT is divergent at all orders as $\tau \rightarrow \pm 0$. Once again in the crossed lines representation  the maximally divergent  contribution of any given diagram is given by the necessary and sufficient condition discussed above. On the other hand this corresponds to select diagrams that are tree-like before averaging and these in turn are generated by the solution of the mean-field equation in presence of the sources $s(x)$.  We have therefore the following
\beq
\langle Q_x(ab)\dots Q_y(cd)\rangle \approx \overline{ q_{x,s}(ab)\dots q_{y,s}(cd)}
\label{cortreespace}
\eeq
where $q_{x,s}(ab)$ is the solution of the mean-field equation {\it in presence} of the source $s(x)$
\beq
0=-\tau-s(x)  -\nabla^2 q_{s,x} -w_1 (q_{s,x}^2)(ab)-w_2 \,q_{s,x}^2(ab)\ .
\label{stochspace}
\eeq
Note that we have not written the term depending on $m_2$ in (\ref{stoch}) since it gives a vanishing contribution because the solution $q_{s,x}$ is an invariant.

In the dynamical case we must remember that the above equation is time-scale invariant and leaves $q_{x,s}(ab)$ undetermined up to a rescaling of time. As discussed previously this invariance is removed by the prescription (\ref{prescr}) that follows from the presence of hidden terms in the action controlling the small-time behavior of the order parameter. Since ${\mathcal L}_s$ is obtained by ${\mathcal L}$ one can easily see that these terms are also present and therefore the very same prescription must be used to determine $q_{x,s}(ab)$.

In the static/replicated case the above stochastic equation is rewritten in terms of its RS solution $q_{s,x}(ab)=q_{s,x}$ where $q_{s,x}$ is the solution of a quadratic stochastic equation:
\beq
0=-\tau-s(x)  -\nabla^2 q_{s,x}+(w_1-w_2) (q_{s,x}^2)\ .
\eeq
This equation does not admit a real solution in finite dimension in the thermodynamic limit, because {\it even} for large positive values of $\tau$ there will be always regions where the local temperature $\tau+s(x)$ is negative.
We have thus recovered diagrammatically the results of \cite{Franz11b}. 

We now turn to the dynamical case. As before we specialize  equation (\ref{stochspace}) to the correlation component of the order parameter:
\beq
\langle C_x(t_a,t_b)\dots C_y(t_c,t_d) \rangle \approx \overline{g_{s,x}(t_a-t_b)\dots g_{s,y}(t_c-t_d)}
\eeq
where $g_{s,y}(t)$ is the solution of the SBR equation 
\beq
 \tau+s(x)=-\nabla^2 g_{s,x}(t) -w_2 \, g_{s,x}^2(t)+w_1{d \over dt}\int_0^t g_{s,x}(t-t')g_{s,x}(t')dt'
\eeq
with the prescription (\ref{prescr}).
At variance with the static case, the function $g_{s,x}(t)$ exists for all values of $s(x)$, therefore the  dynamical theory  is perfectly well defined. 
The above equations define SBR in the finite-dimensional case. We have shown that at each order in the loop expansion the leading divergent terms of the GCT are the same of SBR, thus completing the proof of the results anticipated in \cite{Rizzo14}.
As already noted in the introduction,  only the averaged order parameter was mentioned in \cite{Rizzo14} while it should be clear from the above discussion that the connection with SBR holds as well for {\it all} higher order correlation functions (\ref{genavespace}).

\section{Beyond Perturbation Theory}
\label{beyond}

Let us summarize the results of the previous section.
We started from the GCT action (\ref{action}) that at the mean-field level exhibits a dynamical transition. Loop corrections to the mean-field expressions are all divergent as we approach the transition and we have shown that  the most divergent contribution of each diagram is the same  generated by the loop expansion of SBR.
For later reference we note that this is also what happens for the RFIM or for branched polymers in purely static/replicated context \cite{Parisi79}. 

In general below the upper critical dimension and above the lower critical dimension a system exhibits a critical behavior different from mean-field. However, one can use the perturbative expansion around the mean-field solution to extract the non-mean-field critical exponents by resumming them appropriately.  Note that to apply these resummation techniques one does not need the whole loop expansion but only the {\it leading divergent terms} at each order. Thus, if the leading order divergent terms are the same of a stochastic equation one would argue that the critical behavior is {\it exactly} the same.
In other words the mapping between the effective theory and the stochastic equation holds only at the leading order, {\it i.e.} it is approximate but it is {\it exact} as long as we discuss critical behavior. 
Indeed, in the case of branched polymers one can show that the critical exponents of the two theories are the same \cite{Parisi79}.

In the present context, however, the situation is rather different.
When we consider the GCT beyond the mean-field approximation the critical point is avoided and nothing is divergent. As a consequence, there is no guarantee that: first, we can replace the actual model under study with the effective GCT, and, second, we can replace the GCT with SBR. These kind of problems occur when we are below the lower critical dimensions and are not often considered in the theory of critical phenomena which is focused by definition on what happens between the upper and lower critical dimension.

The first problem is generic to any Landau theory and cannot be discussed within the theory itself.
It depends on the system under study, more precisely its  long-wavelength Hamiltonian must verify two properties: 1) the coefficient $\tau$ of the linear term must vanish for some temperature and 2) higher order terms (quartic and so on) and all other terms in that are present in the full action but not in the GCT can be neglected. 
A sufficient condition for this to happen is that the  system under study exhibits mean-field-like critical behavior in some range of temperatures above the (pseudo)-critical temperature. 
This means that, in that range of temperatures, we can replace the integral of the action (\ref{action}) with its value on the mean-field solution. If this is the case we can describe the critical region, where corrections are instead important, with the GCT. In practice if a given super-cooled liquid exhibits approximately the scaling laws of ideal MCT then one can expect that the GCT provides an accurate description in the crossover region.

The second problem, {\it i.e.} the mapping between the GCT and SBR, is less standard. The zero-dimensional case is the paradigm of the system below the lower critical dimension and therefore it is a good starting point for the discussion.
In the previous section we have shown that there is an {\it exact} mapping between the GCT and a GCT with quenched disorder $s$ on $\tau$:

\begin{displaymath}
{1 \over Z} \int \left(\prod_{(ab)} dQ(ab)\right) Q(ab)\dots Q(cd) \exp[-{\mathcal L}]=
\end{displaymath}
\beq
\overline{ {1 \over Z_s} \int \left(\prod_{(ab)} dQ(ab)\right) Q(ab)\dots Q(cd) \exp[-{\mathcal L_s}]}
\eeq 
where the overline means average over a Gaussian random variable with zero mean and variance $v$. 
The above formula is the result of an Hubbard-Stratonovich integration combined with  $Z_s=Z=1$. We recall that the importance of this last condition should not be underestimated because otherwise $s$ could not be considered an independent (quenched) Gaussian variable.
The result $Z_s=1$ is proven in the appendix at all orders in perturbation theory but it  seems to have a deeper physical origin: the fact that system is at equilibrium and therefore TTI holds.

Now we clearly see  what kind of approximation we are making when we replace the original GCT with SBR: we are treating the action ${\mathcal L}_s$ at the mean-field level {\it i.e.} replacing $Q(ab)$ with the solution of the equation $d{\mathcal L}_s/dQ=0$.
The quality of this approximation depends on the coupling constants of the GCT. In order to proceed let us rescale $s$ to make its variance equal to one. In this case we have 
\bew
\beq
{\mathcal L}_s =  - (\tau+ v^{1/2} s) \int d(ab)Q(ab)-{1 \over
  6}w_1 \int(da\,db\,dc)Q(ab)Q(bc)Q(ca)-{1 \over 6}w_2
\int (da\,db)Q^3(ab)
\eeq
\eew
where we have neglected again the term dependent on $m_2$ because they are not relevant for the present discussion.
Performing the following rescalings:
\beq
Q(ab)=b_Q \tilde{Q}(ab)\ , \ \tau=b_\tau \tilde{\tau}\ , 
\eeq
with 
\beq
b_Q\equiv  v^{1/4}\, w_1^{-1/2} \, , \ b_\tau\equiv  v^{1/2}  \, ,
\eeq
we have: 
\bew
\beq
{\mathcal L}_s =  C\left[- (\tilde{\tau}+ s) \int d(ab)\tilde{Q}(ab)-{1 \over
  6} \int(da\,db\,dc)\tilde{Q}(ab)\tilde{Q}(bc)\tilde{Q}(ca)-{\lambda \over 6}
\int (da\,db)\tilde{Q}^3(ab)\right]
\eeq
\eew
where 
\beq
C=v^{3/4}\,w_1^{-1/2}\ .
\eeq
When ${\mathcal L}_s$ is rewritten in the above form one can easily argue that perturbative corrections to the mean-field approximation are given by an expansion in powers of $O(1/C)$ and therefore the condition for its validity is that the constant $C$ must be large.
Even if $C$ is large,  the random fluctuations can  be neglected if $|\tilde{\tau}| \gg 1$.  This condition defines  the temperature regime where not only ${\mathcal L}_s$ but also the original GCT (${\mathcal L}$) are both well described by the mean-field approximation. 
Summarizing, if the constant $C$ is large  the GCT can be described by the mean-field approximation for large $\tilde{\tau}$, while for  $\tilde{\tau} =O(1)$  the mean-field approximation is wrong. Nevertheless the random action ${\mathcal L}_s$ can still be approximated at the mean-field level, meaning that the crossover is well described by SBR.

Clearly the condition that $C$ is large is violated if the random temperature fluctuations are too small {\it i.e.} $v \ll 1$. In this case the mapping to the stochastic equation is not granted and the crossover is described by the pure cubic theory. Different scalings should be considered, the critical region is still given by $\tilde{\tau}=O(1)$ but we should set $b_\tau= w_1^{1/3}$ and $b_Q=w_1^{-1/3}$.
We note that the condition that $C$ is large is the counterpart for an avoided critical point of  the so-called {\it Harris} criterion that, for a genuine critical point, determines weather or not the disordered and the pure system are in different universality classes \cite{Cardy98}.
Similarly the condition $\tilde{\tau}=O(1)$ should be identified with the so-called {\it Ginzburg} criterion that identifies the region where the mean-field approximation is wrong and takes a different form depending on weather the Harris criterion is verified or not. 

There is at least one case in which the two questions discussed so far can be both answered positively {\it i.e.} the problem of  finite-size corrections to mean-field models.
In this case all coupling constants in the Hamiltonian are proportional to the system-size, say $N$, and it follows that higher order terms not present in the GCT are irrelevant in the crossover region. Furthermore the fact that $v$ and $w_1$ are $O(N)$ implies:
\beq
C=O(N^{1/4})
\label{C0D}
\eeq
meaning that the constant $C$ grows with the system-size and therefore can be taken as large as we want by considering large enough systems. In particular this implies that SBR describes the finite-size corrections to the dynamics of mean-field 1RSB-SG models.
Note that in this case $\tau$ is proportional to $T-T_c$ times a $O(N)$ constant and the crossover region described by SBR is 
\beq
|T-T_c|=O(N^{-1/2})\ ,
\label{TTC0}
\eeq
while ideal MCT scalings hold outside this region.
This scaling is different from what happens in a pure cubic theory where instead $T-T_c=N^{-2/3}$ and it shows that the definition of the Ginzburg criterion depends on weather the Harris criterion is verified or not.

In the present framework one can also discuss the time scale of the $\beta$ regime.
A Landau theory holds as long as  the order parameter is somehow  {\it small}. Clearly this condition is violated for small enough times (when the correlator is approaching the plateau)
and for large enough times (when the correlator leaves the plateau). This allows to rationalize the divergence of the correlator at small and large times as $1/t^a$ and $-t^b$. In both cases the divergences must be interpreted as the correlator respectively entering and exiting the $\beta$ regime where the GCT applies.  Given that the order parameter is $O(b_Q)$ the time scale of the $\beta$ regime can be obtained from $b_Q(t_0/\tau_b)^{-a}=1$ where $t_0$ is the scale of the initial dynamics and it is finite. This leads to  
\beq
\tau_\beta=t_0 b_Q^{-1/a}
\eeq
Note the analogy with (\ref{tauscal}) that can be obtained from the scaling $b_Q=\sqrt{\tau}$ that holds outside the crossover region.
Similarly the (early) $\alpha$ regime can be obtained from $b_Q(\tau_\alpha\tau_\beta)^{-b}=1$ and gives 
\beq
\tau_\alpha=t_0 b_Q^{- 2 \gamma}
\eeq
where $\gamma \equiv 1/(2 a)+1/(2 b)$.
For the models mentioned above whose coupling constants are $O(N)$ we have $b_Q=O(N^{-1/4})$ and thus:
\beq
\tau_\beta=O(N^{ 1 \over 4a})\, , \ \tau_\alpha=O(N^{\gamma/2}) \ .
\eeq
Note that the above scalings describe the increase of the $\alpha$ and $\beta$ time scales when increasing $N$ at fixed $\tilde{\tau}$. The behavior at fixed $N$ and increasing $\tilde{\tau}$ can be discussed in the context of SBR see \cite{Rizzo14,Rizzo15a,Rizzo15b} and displays a satisfactory power-law-to-exponential crossover.

In the following we generalize the above treatment to generic dimension $D$.
In the most general case the constant in front of the space derivative term takes a value $\alpha$:
\beq
{\mathcal L}_{x} \equiv  {\alpha \over 2}  \int d(ab)|\nabla Q_x(ab)|^2+  {\mathcal L}[Q_x] 
\eeq
We need also a rescaling of the space-dependent variable $x$ according to $x=b_x \tilde{x}$ and thus also of the random temperature $s=b_s \tilde{s}$ in order to have $[\tilde{s}(\tilde{x})\tilde{s}(\tilde{y})]=\delta(\tilde{x}-\tilde{y})$.
With the following rescalings:
\beqa
b_{\tau }&=& \alpha ^{\frac{2 D}{D-8}} v^{-\frac{4}{D-8}}
   w_1^{-\frac{D}{D-8}},
\label{btau}   
   \\
   b_Q &=& \alpha ^{\frac{D}{D-8}} v^{-\frac{2}{D-8}}
   w_1^{-\frac{4}{D-8}-1},
   \\
   b_x &=& \alpha ^{-\frac{4}{D-8}} v^{\frac{1}{D-8}}
   w_1^{\frac{2}{D-8}},
   \\
   b_s &=&  \alpha ^{\frac{2 D}{D-8}} v^{\frac{D}{16-2 D}}
   w_1^{-\frac{D}{D-8}},
   \eeqa
we can rewrite the action as
\bew
\beq
{\mathcal L}_s=C \, \int d^D\tilde{x} \left[  {1 \over 2}  \int d(ab)|\nabla \tilde{Q}(ab)|^2- (\tilde{\tau}+ \tilde{s}) \int d(ab)\tilde{Q}(ab)-{1 \over
  6} \int(da\,db\,dc)\tilde{Q}(ab)\tilde{Q}(bc)\tilde{Q}(ca)-{\lambda \over 6}
\int (da\,db)\tilde{Q}^3(ab)\right]\ ,
\eeq
\eew
where the adimensional constant $C$ is given by:
\beq
C =\alpha ^{-\frac{D}{D-8}} v^{\frac{2}{D-8}+1}
   w_1^{\frac{4}{D-8}} \ .
\eeq
The zero dimension discussion can now be repeated along the same lines. If the constant $C$ is large enough the action {\it before} disorder average can be treated at the mean-field level and the mapping to SBR is accurate in the crossover region. For $|\tilde{\tau}| \gg 1$ we are outside the crossover region and also the action {\it after} averaging, {\it i.e.} the  GCT, can be treated at mean-field level. In this region the ideal MCT scalings apply, including an apparent $O(|\tilde{\tau}|^{-1/4})$ divergence of the correlation length that is smeared in the crossover region. The crossover region is identified by the Ginzburg criterion $\tilde{\tau}=O(1)$  where $\tilde{\tau}$ is defined through eq. (\ref{btau}) provided $C$ is large enough \footnote{The Ginzburg criterion can be derived in many ways obtaining different numerical prefactors, but essentially it corresponds to compare $\tau$ with a constant of the same dimension formed with $\alpha$, $v$ and $w_1$. The condition $\tilde{\tau}=O(1)$ can be matched with the Ginzburg criterion for quadratic stochastic equations given as eq. 148 in \cite{Franz12} by recognizing that $m_R\rightarrow (w_1 \tau)^{1/4}$}.

The $C \gg 1$ condition is the analog of the Harris criterion and depends on the specific model considered. Similarly to the zero-dimensional case
one can define {\it ad hoc} models where the two conditions discussed previously hold provided a tunable parameter $M$ is large. One strategy is to increase the number of local microscopic components at the price of loosing quantitative contact with realistic models.
For spin systems a classic way to achieve this is to put $M$ spins on each given lattice point, see \cite{Caltagirone11}. Another strategy is to consider weak long-range interactions, {\it i.e.} taking the so-called Kac limit \cite{Franz04}. 
For supercooled liquids the problem is more complicated (see the discussion in \cite{Mari11}) and at present no well-established tunable model has been proposed \footnote{The model proposed in \cite{Mari11} has been claimed not to display a MCT transition in \cite{Charbonneau14}}. 
The key property of these tunable models is that for {\it large but finite} $M$ the constants $v$, $w_1$, $w_2$ and $\alpha$ are $O(M)$  leading to:
\beq
C=O(M^{2 \over 8-D})
\eeq
that reduces to (\ref{C0D}) for $D=0$.
Similarly to (\ref{TTC0}) the size of the crossover region also shrinks with increasing $M$ as
\beq
|T-T_c|=O(1/M^{4 \over 8-D})
\eeq
The $\alpha$ and $\beta$ time scale behavior for increasing $M$ at fixed $\tilde{\tau}$ can be derived as before and read:
\beq
\tau_\beta=O(M^{1 \over a(4-D/2)})\, , \ \ \tau_\alpha=O(M^{\gamma \over 2-D/4})
\eeq
The corresponding scalings at fixed $M$ and changing $\tilde{\tau}$ can be derived in the context of SBR \cite{Rizzo15b}. We only quote the result for $\tau_\alpha$ respectively for $\tilde{\tau} \rightarrow -\infty$ and for $\tilde{\tau} \rightarrow \infty$
\beq
\tau_\alpha=O(|\tilde{\tau}|^{-\gamma})\, , \ \ \ln \tau_\alpha=O(\tilde{\tau}^{2-D/2})
\eeq

Previously we noticed that if a system exhibit MCT-like scalings in some range of temperatures then it is natural to expect that the crossover region is described by the GCT.
Now we want to discuss a practical criterion in order to determine weather the GCT should be described by SBR in the crossover region. Approaching the crossover region from above one expects that fluctuations (the full propagator) are well described by the bare propagator. The main feature of the bare propagator is the presence of the so-called double pole term proportional to $\chi(ab)\chi(cd)$ which leads to larger dynamical fluctuations with respect to those associated to temperature variations ($\chi(ab)$). On the other hand the double-pole term would have a  subdominant prefactor if the constant $C$ is small and therefore the fact that one observes fluctuations proportional to $\chi^2$ (as reported in some notable case for supercooled liquids \cite{Biroli06}) can be taken as an indication that the constant $C$ is large and that the crossover region may be accurately described by SBR.

We conclude this section noticing that the above scalings with $M$ are singular at $D=8$ and make no sense for $D>8$. They were obtained under the implicit assumption that the GCT and SBR admit a canonical continuous limit in a field-theoretical sense and the ill-defined scalings for $D>8$ must be taken as an indication that the continuous limit is not well defined for $D>8$.  As discussed in the conclusions, this apparently exotic feature has important physical quantities, pointing to a different nature of the crossover above and below $D=8$.

\section{Conclusions}
\label{Conclusions}

The starting point of our discussion is the fact that 1RSB-SG and MCT display a dynamical transition with the same features. In particular, the equation for the critical correlator is the same.
The transition is second-order in nature, {\it i.e.}, it is characterized by a diverging correlation length, a classic result in SG that can be also established within MCT through IMCT. In both cases the transition is mean-field in nature, in the sense that the order parameter obeys an equation with coupling constants that are regular functions of the external parameters ( {\it e.g.} temperature or density). This is equivalent to the assumption that the (Dynamical) Gibbs Free energy ({\it i.e.} the integral of the action) has also regular coupling constants. The theory of phase transitions tells us that this result has to be put under scrutiny and notably fails below the upper critical dimension. Indeed, physics requires that the microscopic Hamiltonian ({\it i.e.} the action) has a regular dependence on the external parameter, {\it not} 
 the Gibbs free energy ({\it i.e.} the integral of the action). The discussion in section \ref{beyond} allows to make this point explicit: for $|\tilde{\tau}|\gg 1$ one can take the saddle point in the integral and trade the Hamiltonian  for the Gibbs free energy, but in the critical region $|\tilde{\tau}|=O(1)$ this is not correct.
On the other hand in the critical region one can study an effective action that is a simple function (a polynomial) of the order parameter on long wavelength, {\it i.e.} a Landau theory.

In order to choose this effective theory we can first consider the so-called FM limit, in which all dynamical correlation functions take a simplified (RS-like) structure. It follows that in the FM limit the effective action must also display a RS-like form, indeed, in this way one can show that equilibrium dynamics becomes equivalent to a replicated problem with $n=1$ replicas.
The replicated problem was originally studied in \cite{Franz11b} and a mapping to a static stochastic equation was discovered. The stochastic static equation, however, is ill-defined suggesting that the arrested state is unstable and that the transition is in fact avoided, as it is expected from the physics of supercooled liquids.
Therefore in order to have a well-defined theory of the crossover one must abandon the FM limit and the static replica approach (that is only well defined in presence of a stable glassy phase).

Outside the fast motion limit, the dynamical correlations do not have the simplified RS-like structure and correspondingly the dynamical action contains additional terms depending explicitly on the microscopic dynamical scale $\Gamma_0$.
An important technical point  is that the FM/RS mean-field equations admit also non-FM dynamical solutions. At variance with the static case, these solutions exist both above and below the critical temperature and are precisely the same of the critical correlator in MCT.  The absence of $\Gamma_0$-dependent terms in the equations induces a spurious time-scale invariance that can be removed invoking matching arguments with the physics on microscopic times. Formally we supplement the dynamical action with a prescription to remove time-scale invariance that is justified by the presence of the hidden $\Gamma_0$-dependent terms.

The identification of the correct effective theory for a given problem can be done in some cases explicitly, but in general it is more satisfactory to invoke some symmetry principle. 
For 1RSB-SG one can explicitly show that the effective theory is the GCT given by eq. (\ref{action}). For MCT this is not possible, because we do not know how to compute within  the theory many quantities, notably four-point correlation functions. On the other hand, time-scale invariance provides us a principle to carry on the identification: ideal MCT is time-scale invariant and if we require that the effective theory preserves this feature we are naturally led to the glassy critical theory.

The main part of the paper was devoted to the study of the GCT.
Given that, as usual, an exact computation of the averages of the order parameter is not possible, we first resorted to a diagrammatic loop expansion.
Let us recall that, if $M$ is some large parameter and the coupling constants  $\tau$, $m_2$, $m_3$, $w_1$ and $w_2$ are $O(M)$ (a special situation we discussed thoroughly in section (\ref{beyond}), the loop expansion corresponds to an expansion in powers of $1/M$. As usual the coefficients of the various powers of $1/M$ are divergent as $\tau \rightarrow 0$ and we have seen that, at any given order $1/M$, the leading divergent term is the same that is generated within a similar loop expansion of SBR.
Thus I have provided the proof of the result that was anticipated in \cite{Rizzo14} and further discussed in \cite{Rizzo15a}. Note that in that paper only the average two-point function was discussed while we have shown here that SBR provides a description of {\it higher-order correlations} as well, {\it e.g.} the four-point function.   

In general if a second-order phase transition is not washed out by fluctuations one can appropriately re-sum the divergent terms and obtain information on the non-mean-field critical exponents. In these cases a mapping with a stochastic equation would imply that the critical exponents of the original theory are the same of the stochastic equation. In the present case, however, the transition is avoided and we are more in the situation that occurs below the lower critical dimension, therefore the mapping between the GCT and SBR is only approximate.
Furthermore, the avoided nature of the transition within SBR, as discussed in \cite{Rizzo14,Rizzo15a,Rizzo15b}, is clearly due to non-perturbative effects. Therefore one would like to be sure that the same non-perturbative effects are at work in the GCT, thus establishing a connection between the two theories beyond perturbation theory. 
This problem has been discussed in the last section. It turned out that one can establish an {\it exact} mapping between the GCT and a theory with {\it quenched} disorder. This allows to understand clearly the approximation involved in replacing the GCT with SBR:
it amounts to make a mean-field approximation separately on each instance {\it after} the quenched disorder has been generated.
In formulas the exact mapping corresponds to eq. (\ref{corexa}) while the SBR approximation corresponds to eq.  (\ref{cortree}).

Note that the exact mapping with a quenched problem was not discussed in \cite{Rizzo14} and it is more significant than the mapping with SBR: the latter holds only for the {\it leading} divergent term at each order in the loop expansion. 
Furthermore the mapping allows a compact quantitative and non-perturbative discussion of the range of the coupling constants where the mapping to SBR is valid.  Besides, even if for some values  of the coupling constants SBR cannot be applicable quantitatively, qualitatively the idea that dynamical arrest is washed away by random fluctuations of the temperature may remain valid.

The exact mapping implies that one can drop the perturbative loop expansion of the second section altogether and study only the quenched theory and its relation with SBR. I have chosen not to do so because the diagrammatic expansion is very instructive but also because the mapping with the quenched problem relies on the property $Z=Z_s=1$ that were proven at all orders in the appendix.
 Much as the crucial property of invariant matrices (\ref{zerocon}), they appear to be a consequence of the system being at equilibrium and thus could likely be proven non-perturbatively as well.
 
As an additional remark note that while the static replica treatment is not able to describe the dynamical crossover, it is instead applicable to problems where there is a genuine second-order phase transition at the mean-field level, {\it i.e.} pinned systems \cite{Franz13a,Biroli14}. In the replica case the condition $Z_s=1$ is almost trivial and allows to establish in a definitive way a mapping between pinned systems and systems with quenched random fields that have been advocated often in recent literature \cite{Cammarota12,Franz13a,Franz13b,Biroli14}.

The discovery of \cite{Franz11b} that the static/replicated action leads to an ill-defined static stochastic equation opened the quest for a well-defined theory. 
Given that the static/replicated problem is equivalent to a spinodal in the RFIM, one could think that their solution is also the same. For instance the authors of ref. \cite{Nandi16} recently claimed that: \textquotedblleft The theoretical understanding of
the spinodal of the RFIM in finite dimensions is therefore
the crucial missing step to conclusively assess the physical
content of the MCT and the status of the dynamical
transition predicted by the MF theory of glass-forming
liquids\textquotedblright. Furthermore, the argued in favour of the survival of the RFIM spinodal  beyond mean-field, pointing to a discrepancy with SBR where the transition instead is avoided.

Both results are likely correct,  the problem is which one is relevant to fix the ill-defined static theory associated to MCT. Now that the complete derivation has been presented, it should be clear that SBR provides the sole correct answer.  This is because I extended the very same derivation of \cite{Franz11b} from a static to a dynamical context and this was enough to obtain a well defined theory. One can now see that the analogy between the MCT crossover and the RFIM spinodal holds at the level of the statics but {\it not} at the level of the dynamics.
The essential difference is that introducing dynamics in the field theory is not helpful  in the RFIM spinodal problem.
Let us consider again the zero-dimensional case. By a standard computation one can show that the problem is equivalent to a dynamical stochastic equation of the following form:
\beq
{  d\over dt }\,g_{s}(t) =\tau+s-w\, g_{s}^2(t)\ ,
\eeq 
to be compared with the SBR expression 
\beq
w_1\left( - g_{s}^2(t)+{d \over dt}\int_0^t g_{s}(t-t')g_{s}(t')dt' \right) =\tau+s-w\, g_{s}^2(t)\ ,
\eeq 
where $w=w_1-w_2>0$. The difference between the two equations is the l.h.s. that governs the dynamics and it is absent in a purely static context.
For negative $\tau+s$ the solution of SBR goes to $-\infty$ at large times as $-t^b$ and it is therefore defined at all times, this implies that the average over solutions  $\overline{g_s(t)}$ is also well-defined.
 Instead the solution of the spinodal equation goes to $-\infty$ as $-1/(t_L-t)$ in a  finite time $t_L \propto 1/|\tau+s|^{1/2}$.
As a consequence the average over solutions $\overline{g_s(t)}$ is still ill-defined for the spinodal RFIM problem because $|\tau+s|$ in the average can be as large as we want yielding a contribution that diverges before any finite time. One can argue that the problem is not fixed by the presence of the gradient term and remains also in finite dimension. 
Of course the problem of the RFIM spinodal must have its own solution, although at present not fully understood. The point is that the corresponding problem in MCT has already its solution and it is provided by SBR.

Given the analogy between the replicated problem and the RFIM spinodal of \cite{Franz11b}  one can easily derive a Ginzburg criterion for the perturbative expansion leading to an upper critical dimension $D_u=8$.  Both the notion of an MCT universality class and that of an upper critical dimension $D_u=8$ have been often employed in subsequent discussions of MCT criticality. However one should not omit that they are meaningless concepts in a purely static context given that the corresponding stochastic equation is ill-defined. 
As discussed in the last section, SBR allows instead to give a precise meaning to these notions.

For a genuine second-order phase transitions, universality means that the critical region can be described by a unique effective field theory that does not depend on the microscopic details of the model.  Below the upper critical dimension this theory is the renormalized theory. For $D>D_u$ the renormalized theory does not exist, meaning that the continuum limit of the field theory is not defined. Still the critical region is described by a unique model-independent theory, which is nothing but the Gaussian theory.

In the case of MCT there is no upper critical dimension because there is no transition, still $D=8$ plays a special role.  In the last section we saw that for $D<8$ all coupling constants in the GCT except $\lambda$ and $C$ can be factored out. Then, assuming that $C$ is large, one can replace the GCT with SBR, obtaining an object that depends only on $\lambda$.
Therefore there is a kind of {\it weak} universality:  the effective theory of the crossover depends on the original model solely through a single non-universal constant $\lambda$. Instead, a genuine second-order phase transition has no model-dependence at all, because the coupling constant has a universal value.

For $D>8$ we argued that the continuum limit of SBR does not exist. Again this is similar to what happens for a true second-order phase transition. 
In the last case, however, the Gaussian theory describes the critical point and the critical exponents take their mean-field values.  
Instead, for MCT, there is no such thing as a Gaussian/mean-field theory of the crossover:
by definition the crossover is the region where mean-field behavior is violated!
This implies that the nature of the crossover must be qualitatively different above and below $D=8$, SBR applies for $D<8$  while the $D>8$ case is an open problem.
Actually I expect that for $D>8$  the system is more \textquotedblleft mean-field\textquotedblright  in the sense that the crossover region is smaller. This notion can be made more precise for tunable models for values of the parameter $M$ large but not infinite. In the last section we saw that the crossover region decreases as $M^{-{4 \over 8-D}}$ in the large-$M$ limit. This scaling-law must break down for $D>8$ where we expect a smaller value, most likely exponentially small in $M$.

Recently, it has been shown that hard-spheres models display an ideal MCT transition in the infinite-dimension limit $D \rightarrow \infty$ \cite{Kurchan12}.
In principle this opens the way to an expansion in $1/D$, although at present a characterization of the crossover in large but finite dimension is lacking. The previous considerations, however, suggest that such an expansion would be useless down to physical dimension because the nature of the crossover changes at $D=8$.

I conclude noticing that tunable models are also important because they guarantee that both  
 the GCT and SBR are fully consistent theories from a physical point of view, included their less intuitive properties.

{\em Acknowledgments.} ~~ I thank L. Leuzzi for a careful reading of the manuscript. 
I have used towards development of this project  
funding from the European Research Council
(ERC) under the European Union’s Horizon 2020 research and innovation
programme (grant agreement No [694925]).

\appendix
\section{Invariant Matrices}
\bew

In the dynamical treatment the order parameter has the structure of a matrix $A(b)$
\beq
A(ab)=C_A(t_a,t_b)+\tilde{R}_A(t_a,t_b)n_b+\tilde{R}_A(t_b,t_a)n_a+X_A(t_a,t_b)n_a n_b\ .
\eeq
Thus the matrix $A(ab)$ is parameterized by three functions $C$,$\tilde{R}$ and $X$ and the symmetry $A(ab)=A(ba)$ implies that $C(t,t')=C(t',t)$ and $X(t,t')=X(t',t)$, the function $\tilde{R}(t,t')$ instead is not symmetric. 
When we integrate an action  of the form (\ref{action}) over the order parameter all components $C$, $\tilde{R}$ and $X$ to be integrated upon are  independent, however an essential role is played by the notion of {\it invariant} in which the matrix is parameterized in term of a single function $C_A(s)$ defined for $s\geq 0$:
\beq
A(ab) \rightarrow C_A(s)
\eeq
according to:
\beq
C_A(t,t')=C_A(|t-t'|)
\eeq
\beq
\tilde{R}_A(t,t')=\tilde{R}_A(t-t')=\theta(t-t') {d \over d t'}C_A(t-t') 
\eeq
\beq
X_A(t,t')=0
\eeq

Two important properties that we have used extensively in the paper concern products of invariant matrices.
The first one is that given two invariant matrices $A$ and $B$  the product $A(ab)B(ab)$ is also an invariant parameterized by a function $C_A(t)C_B(t)$
\beq
A(ab)B(ab) \rightarrow C_A(s)C_B(s)
\eeq
The second property is that the matrix product $\int A(ac)B(cb)dc$ with respect to the integral $\int A(c)dc \equiv \int_{0^+} A(t_c,\eta_c) dt_c d \eta_c+A(0,0)$ is also an invariant (and therefore invariant matrices commute). A straightforward computation shows that the function $C_{A\cdot B}(t)$ that parameterizes the dot product takes the form:
\beq
\int A(ac)B(cb)dc \rightarrow C_{A \cdot B}(s)=C_A(0)C_B(s)+C_A(s)C_B(0)-{d \over ds}\int_0^s C_A(s-y)C_B(y)dy
\eeq
The above expression has been used in subsection (\ref{mfsol}) to show that the mean-field equation gives the equation for the critical correlator of MCT. 
One can easily verify that $\int A(ab)da$ does not depend on $b$ for an invariant matrix, indeed we have:
\beq
\int A(ab)\, da=\int_0^{t_b} \tilde{R}_A(t_b,t_a)dt_a+C_A(t_b,0)=-\int_0^{t_b} { dC_A\over ds }ds+C_A(t_b)=C_A(0)
\eeq
In the case in which the invariant matrix satisfies also $A(aa)=0 \rightarrow C_A(0)=0$ we obtain the essential relationship (\ref{zerocon}).
Note that the property of having $C_A(0)=0$ is clearly stable under the product.
It is also invariant under the dot product indeed we have:
\beq
C_{A \cdot B}(0)=C_A(0)C_B(0)
\eeq
If the limits $C_A(\infty)$ and $C_B(\infty)$ exist we have:
\beq
C_{A \cdot B}(\infty)=C_A(0)C_B(\infty)+C_B(0)C_A(\infty)-C_B(\infty)C_A(\infty)
\eeq
and one can verify the matching with the corresponding expressions for RS matrices in the limit $n \rightarrow 1$
\beq
(A B)_d=a_d b_b+(n-1) a_{off} b_{off}\, , \  (A B)_{off}=a_d b_{off}+b_d a_{off}+(n-2) a_{off} b_{off}
\eeq
where the subscript $d$ and $off$ refer to the diagonal and off-diagonal elements of the RS matrices. Thus we see that the stability of the zero-diagonal condition is specific of the formalism, it does not hold for $n \neq 1$ and for off-equilibrium dynamics.

\subsection{Generalized Invariants}

The notion of invariant matrix can be generalized to objects that depend on many indexes:
\beq
A(abc \dots m)
\eeq
In the replica case a generalized invariant  must be invariant under a permutation of the indexes 
\beq
A(\pi(a) \pi(b) \pi(c) \dots \pi(m))=A(abc \dots m)
\eeq
In the dynamical case we define an invariant by the following properties:
i) It must be time-translational invariant;

ii) the $\eta$ component associated to the maximum time must vanish, {\it i.e.} if, say $t_a=\max(t_a,t_b,t_c \dots t_d,t_e)$ and we introduce the following notation for the scalar and $\eta$ components:
\beq
A(a,b,c,\dots ,d,e)=A_{s_a}(t_a,b,c,\dots,d,e)+A_{\eta_a}(t_a,b,c,\dots,d,e)\eta_a
\eeq
we have:
\beq
A_{\eta_a}(t_a,b,c,\dots,d,e)=0\ .
\eeq
This condition enforces {\it causality}: a correlation function is unaffected by a perturbation at later times.

iii) if, say, $t_e$ is the minimum time {\it i.e.}  $t_e=\min(t_a,t_b,t_c \dots t_d,t_e)$ then with the notations above we have:
\beq
A_{\eta_e}(a,b,t_c,\dots,d,t_e)={d \over dt_e}A_{s_e}(a,b,c,\dots,d,t_e)
\label{mincond}
\eeq
Therefore depending on the ordering of the time $\{t_a,t_b,t_c,\dots,t_d,t_e\}$ we have the two relations above for the components of the maximum and of the minimum.
Note that condition ii) is consistent with {\it causality}: a correlation function is unaffected by a perturbation at later times and condition iii) is consistent with FDT. Thus it is natural to expect that generic equilibrium averages have the structure of generalized invariants.

It can be checked that the product of two invariants is also an invariant in the sense that the three properties above are verified. This include that case in which some of the indexes are the same:
\beq
A(abc \dots de)B(a'bc \dots de)
\eeq
Another important property is that the integral  of an invariant
\beq
\int de A(abc\dots de)
\eeq
is still an invariant. 
In order to prove this statement let us assume without loss of generality that when the index $e$ is integrated out $t_a$ is the largest time and $t_d$ is the smallest time.
If we write:
\beq
A(a,b,c,\dots ,d,e)=A_{s_e}(a,b,c,\dots,d,t_e)+A_{\eta_e}(a,b,c,\dots,d,t_e)\eta_e
\eeq
we have that 
\beq
\int de A(abc\dots de)=A_{s_e}(a,b,c,\dots,d,0)+\int dt_e A_{\eta_e}(a,b,c,\dots,d,t_e)\ .
\eeq
Applying the properties $ii)$ and $iii)$ we have:
\beq
\int de A(abc\dots de)=A_{s_e}(a,b,c,\dots,d,t_d)+\int_{t_d}^{t_a} dt_e A_{\eta_e}(a,b,c,\dots,d,t_e)\ .
\label{simpli}
\eeq
from the above formula we can easily check that the l.h.s remains TTI after the integration.
We also see that in the above expression, $t_a$ is always the maximum time therefore the l.h.s.  satisfies also the causality property ii).
In order to verify the minimum condition we write explicitly the scalar and $\eta$ components with respect to variable $d$.
Now  the $\eta_d$ component of (\ref{simpli}) reads:
\beq
A_{\eta_d s_e}(a,b,c,\dots,t_d,t_d)+\int_{t_d}^{t_{a}} dt_e A_{\eta_d \eta_e}(a,b,c,\dots,t_d,t_e)\ .
\label{ch0}
\eeq
 The derivative of the scalar component of (\ref{simpli}) is made of two terms, the first is:
\beq
{d \over dt_d}A_{s_e s_d}(a,b,c,\dots,t_d,t_d)=A_{\eta_d s_e}(a,b,c,\dots,t_d,t_d)+A_{\eta_e s_d}(a,b,c,\dots,t_d,t_d)
\label{ch1}
\eeq
where we have used property ii) of $A$.
The second is:
\beqa
{d \over dt_d}\int_{t_d}^{t_{a}} dt_e A_{s_d\eta_e}(a,b,c,\dots,t_d,t_e) & = &-A_{s_d\eta_e}(a,b,c,\dots,t_d,t_d)+\int_{t_d}^{t_{a}} dt_e {d \over dt_d} A_{s_d\eta_e}(a,b,c,\dots,t_d,t_e)=
\nonumber
\\
& = &-A_{s_d\eta_e}(a,b,c,\dots,t_d,t_d)+\int_{t_d}^{t_{a}} dt_e  A_{\eta_d\eta_e}(a,b,c,\dots,t_d,t_e)\ .
\label{ch2}
\eeqa
where the last equality follows from the fact that inside the integral $t_d$ is the minimum and eq. (\ref{mincond}) applies.
We see that the sum of (\ref{ch1}) and (\ref{ch2}) is equal to (\ref{ch0}), thus proving that the $\int de A(abcd\dots de)$ verifies the minimum condition (\ref{mincond}).

It is clear that higher order equilibrium correlation functions are associated to generalized invariants much as the two-time correlations are associated to invariant matrices  defined by (\ref{2time}). Since the generic averages (\ref{genave}) are associated to equilibrium correlation functions they must be generalized invariants, this indeed can be verified self-consistently at every order in the expansion around the mean-field solution.

\eew
  
\section{The partition function}

In this section we want to show that the partition function $Z_s$ and $Z$ are both equal to one exactly.
In order to do so we will show that $\ln Z_s$ is zero on the mean-field solution $q_s(ab)$ and that corrections to the mean-field  values vanish at all orders.

Every term in the actions ${\mathcal L}$ and ${\mathcal L}_s$ has a RS-like structure. 
The general RS-like term is formed considering a product of $Q$'s where some indexes are equal and integrating over all indexes. According to the previous subsection it follows that such a procedure leads to a generalized invariant, more specifically if we consider one $Q(ab)$ in the product and integrate over all the remaining indexes we will end up with an expression of the form $\int da db q(ab) E(ab)$ where $E(ab)$ is the result of all the other integrations and it is thus an invariant. Since $q(aa)=0$, it follows that $q(aa)E(aa)=0$ and therefore the condition (\ref{zerocon}) implies that the result is equal to zero when evaluated over an invariant $Q(ab)$.
This implies that the result 
\beq
{\mathcal L}[q(ab)]=0
\eeq
holds for any RS-like action, not only for the cubic actions considered here and also for any invariant $q(ab)$, not only for the solution of the mean-field equation.
The above result is trivial in the replica case, it follows from the fact that every RS term is proportional to $n-1$ when evaluated on a RS matrix.

Neglecting irrelevant constants the Gaussian correction to the mean-field result $\ln Z=0$ is given by the trace of the logarithm of the bare propagator, $\mathrm{Tr} \ln G$.
We recall that $G$ is a symmetric matrix on the vector space of symmetric zero-diagonal matrices therefore $\ln G$ is also a matrix on this vector space.
Now it is important to realize that {\it the bare propagator $G$ is a generalized invariant}  as a function of its four indexes. We know that it should be so because it is associated to an equilibrium four-time correlation, but one can also prove it self-consistently within the theory noticing that it is the inverse of the mass matrix that, being obtained from a RS-like action,  is itself an invariant.
It follows that $\ln G$ is also a generalized invariant. 
On the other hand the trace of a linear operator $A(ab)(cd)$ is given by $\int d(ab) A(ab)(ab)$, therefore if $A(ab)(cd)$ is an invariant it follows that $A(ab)(ab)$ is also an invariant and thus  the trace vanish. This is also true for the identity operator for which one has
\beq
\mathrm{Tr} I=\int d(ab)=0
\eeq
according to eq. (\ref{zerocon}). It follows that the Gaussian correction is also exactly zero for a generic RS-like action:
\beq
{1 \over 2}\mathrm{Tr} \ln G=0
\eeq
Again in the replica case this result is trivial because the trace of a RS operator $M(ab)(cd)$ is proportional to $n(n-1)$.

Higher order corrections to $\ln Z$ are given by all connected vacuum (no external legs) diagrams, like the ones shown in the first line of fig. (\ref{f1}). 
Given one of these diagrams  we have to multiply a propagator $G(ab)(cd)$ for each leg and then perform the integral over the indexes at the vertex. Since we have $G(aa)(cd)=0$ the integrals can be made on unrestricted indexes.
We are integrating a product of generalized invariants, thus the result, before and after each integration, is also a generalized invariant.  If we fix a couple of indexes, say $ab$, on one internal line $G(ab)(cd)$ and integrate over all other indexes the result $E(ab)$ will be an invariant with zero diagonal, therefore it will give a vanishing contribution after the final integration over $da db$ according to (\ref{zerocon}). 
It is clear that this result holds because the bare propagator is invariant and the vertexes are RS-like and therefore it holds also for the generic action containing all possible RS-like vertices.
This shows that all vacuum diagrams generated by a RS-like action give a vanishing contribution and complete the proof that $Z=Z_s=1$ exactly. Note that although we have derived this property diagrammatically it appears to be intrinsically related to the system being at equilibrium.
By similar arguments one can show that the generic averages (\ref{genave}) generated by an RS-like action expanded near the dynamical mean-field solution are generalized invariants as expected on physical grounds. 

The previous arguments can be also used to understand the necessary and sufficient condition for a non-vanishing diagram discussed in subsection \ref{diagram}. In the cross-line representation each crossed line gives a factorized contribution and can be opened and replaced with a source $s$ that shifts $\tau$ by an amount $\sqrt{v}$. Therefore one is left with the evaluation of some disconnected diagrams. According to the discussion of the previous paragraph if one of the disconnected diagrams has no external lines it is a diagram that contribute to $\ln Z_s$ and therefore vanishes. This implies that in the crossed line representation the necessary and sufficient condition for a graph to give a non-zero contribution is that once all crossed lines are opened each diagram of full lines contains an external line. 

In principle the action (\ref{action}) may contain additional cubic RS-like terms \cite{Temesvari02}. The presence of these RS-like terms therefore does not alter the exact mapping between the GCT and the GCT with quenched disorder.
Furthermore at the level of the mean-field equation these terms do not play any role, as the terms proportional to $m_2$ and $m_3$, and therefore they do not show up in SBR.

\end{document}